\documentclass[aps, superscriptaddress]{revtex4-1}

\pdfoutput=1
\usepackage{graphicx}
\usepackage{amsmath}
\usepackage{amssymb}
\usepackage{bm}
\usepackage{color}

\newcommand{\pa}{\partial}
\newcommand{\la}{\left\langle}
\newcommand{\ra}{\right\rangle}
\newcommand{\mean}[1]{\langle{#1}\rangle}

\newcommand{\Tr}{{\rm Tr}\hspace{0.07cm}}

\newcommand{\argmax}{\mathop{\rm arg~max}\limits}

\definecolor{darkgreen}{rgb}{0,0.4,0}

\begin{document}
\title{Semiclassical Phase Reduction Theory for Quantum Synchronization}
\author{Yuzuru Kato}
\email{Corresponding author: kato.y.bg@m.titech.ac.jp}
\affiliation{Department of Systems and Control Engineering,
	Tokyo Institute of Technology, Tokyo 152-8552, Japan}

\author{Naoki Yamamoto}
\affiliation{
	Department of Applied Physics and Physico-Informatics, 
	Keio University, 
	Kanagawa 223-8522, Japan}
	
\author{Hiroya Nakao}
\affiliation{Department of Systems and Control Engineering,
	Tokyo Institute of Technology, Tokyo 152-8552, Japan}
\date{\today}

\begin{abstract}
We develop a general theoretical framework of semiclassical phase reduction for analyzing synchronization of quantum limit-cycle oscillators.
The dynamics of quantum dissipative systems exhibiting limit-cycle oscillations are reduced to a simple, 
one-dimensional classical stochastic differential equation approximately describing the phase dynamics of the system under the semiclassical approximation.
The density matrix and power spectrum of the original quantum system can be approximately reconstructed from the reduced phase equation. 
The developed framework enables us to analyze synchronization dynamics of quantum limit-cycle oscillators using the standard methods for classical limit-cycle oscillators in a quantitative way.
As an example, we analyze synchronization of a quantum van der Pol oscillator under harmonic driving and squeezing, including the case that the squeezing is strong and the oscillation is asymmetric.
The developed framework provides insights into the relation between quantum and classical synchronization and will facilitate systematic analysis and control of quantum nonlinear oscillators.
\end{abstract}

\maketitle

%%%%%%%%%%%%%%%%%
%%% Seciton 1 %%%
%%%%%%%%%%%%%%%%%

\section{Introduction}

Spontaneous rhythmic oscillations and synchronization arise
in various science and technology fields,
such as laser oscillations, electronic oscillators, and spiking neurons
\cite{winfree2001geometry, kuramoto1984chemical, 
	ermentrout2010mathematical, pikovsky2001synchronization,  
	glass1988clocks, strogatz1994nonlinear}.
Various nonlinear dissipative systems
exhibiting rhythmic dynamics can be modeled as
limit-cycle oscillators. A standard theoretical framework for 
analyzing limit-cycle oscillators 
in classical dissipative systems is the {\it phase reduction theory}~\cite{winfree2001geometry, kuramoto1984chemical, 
	nakao2016phase, 
	ermentrout2010mathematical,ermentrout1996type,brown2004phase}.
By using this framework, we can systematically reduce 
multi-dimensional nonlinear dynamical equations
describing weakly-perturbed limit-cycle oscillators
to a one-dimensional phase equation
that approximately describes the oscillator dynamics.
The simple semi-linear form of the phase equation, characterized only by the {\it natural frequency} and {\it phase sensitivity function} (PSF) of the oscillator, 
facilitates detailed theoretical analysis of the oscillator dynamics.

The phase reduction theory has been successfully used to analyze universal properties of limit-cycle oscillators in a systematic way, such as synchronization of oscillators with periodic forcing and mutual synchronization of coupled oscillators~\cite{winfree2001geometry, kuramoto1984chemical, 
	ermentrout2010mathematical, pikovsky2001synchronization,  
	glass1988clocks, strogatz1994nonlinear}.
It has been essential in the understanding of synchronization phenomena in classical rhythmic systems, for example, the collective synchronization transition of a population of oscillators and oscillatory pattern dynamics in spatially extended chemical or biological systems~\cite{winfree2001geometry, kuramoto1984chemical}.
Recently, generalizations of the phase reduction theory to non-conventional physical systems, such as time-delayed oscillators~\cite{kotani2012adjoint,novivcenko2012phase}, piecewise-smooth oscillators~\cite{shirasaka2017phase2}, collectively oscillating networks~\cite{nakao2018phase}, and rhythmic spatiotemporal patterns~\cite{kawamura2013collective,nakao2014phase}, have also been discussed.

Recent progress in experimental studies
has revealed that synchronization
can take place in coupled nonlinear oscillators
with intrinsically quantum-mechanical origins,
such as micro and nanomechanical oscillators
\cite{shim2007synchronized,zhang2012synchronization,
	*zhang2015synchronization,bagheri2013photonic, matheny2014phase, matheny2019exotic},
spin torque oscillators \cite{kaka2005mutual},
and cooled atomic ensembles \cite{weiner2017phase, heimonen2018synchronization}.
Moreover, theoretical studies have been performed on
the synchronization of nonlinear oscillators
which explicitly show quantum signatures
\cite{
	ludwig2013quantum,weiss2016noise,
	amitai2017synchronization,
	xu2014synchronization,xu2015conditional,
	lee2013quantum,lee2014entanglement,hush2015spin,
	roulet2018synchronizing,*roulet2018quantum,*koppenhofer2019optimal,
	nigg2018observing,
	lee2013quantum,
	walter2014quantum,
	sonar2018squeezing,
	lee2014entanglement,walter2015quantum,
	lorch2016genuine,
	ishibashi2017oscillation, amitai2018quantum,
	navarrete2017general, weiss2017quantum,
	hriscu2013quantum, hamerly2015optical,
	lee2013quantum2,mari2013measures,ameri2015mutual,
	witthaut2017classical,
	davis2018dynamics,
	lorch2017quantum,
	de2014synchronization},
such as optomechanical oscillators
\cite{ludwig2013quantum, weiss2016noise, amitai2017synchronization},
cooled atomic ensembles \cite{xu2014synchronization, xu2015conditional},
trapped ions \cite{lee2013quantum, lee2014entanglement, hush2015spin},
spins \cite{roulet2018synchronizing, *roulet2018quantum},
and superconducting circuits \cite{nigg2018observing}. 
In particular, a number of studies have analyzed the quantum van der Pol (vdP) oscillator~\cite{lee2013quantum},
which is a typical model of quantum self-sustained oscillators, 
for example, synchronization of a quantum vdP oscillator by harmonic driving~\cite{walter2014quantum, amitai2017synchronization} or squeezing \cite{sonar2018squeezing},
mutual synchronization of coupled quantum vdP oscillators~\cite{lee2014entanglement,walter2015quantum},
and quantum fluctuations around oscillating and locked states of a quantum vdP oscillator~\cite{navarrete2017general, weiss2017quantum}.

In addition to its fundamental importance as a novel physical phenomenon 
where nonlinear and quantum phenomena have combined effect, 
quantum synchronization may also be useful in developing  
metrological applications, such as the improvement of the measurement accuracy in 
the Ramsey spectroscopy for atomic clocks \cite{xu2015conditional}
and the precise measurement of the resistance standard with a superconducting device \cite{hriscu2013quantum}; an application of the limit-cycle oscillation to analog memory in a quantum optical device \cite{hamerly2015optical} has also been considered.

\begin{figure} [!t]
	\begin{center}
		\includegraphics[width=0.6\hsize,clip]{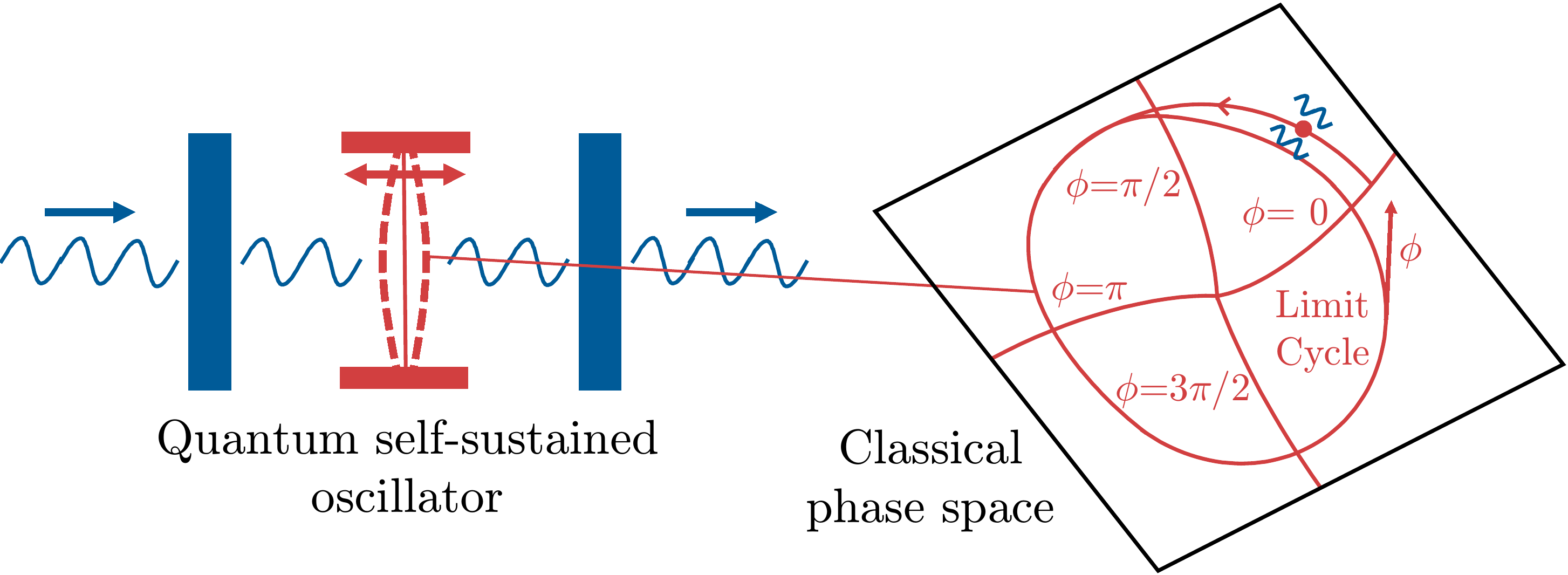}
		\caption{
			A schematic diagram of the semiclassical phase reduction for
			quantum synchronization. A quantum self-sustained oscillator,
			which has a stable limit-cycle solution in the classical limit, 
			can be described by an approximate one-dimensional stochastic differential
			equation for a phase variable $\phi$ that characterizes the system state.
			The system state can be approximately reconstructed from the reduced phase equation.
		}
		\label{fig_1}
	\end{center}
\end{figure}

Considering the importance of phase reduction for analyzing synchronization of 
classical nonlinear oscillators, we aim to develop a phase reduction theory also for quantum nonlinear oscillators.
In the analysis of quantum synchronization, phase-space approaches using the quasiprobability distributions of quantum systems are commonly employed. 
In a pioneering study, Hamerly and Mabuchi~\cite{hamerly2015optical} derived a phase equation from the stochastic differential equation (SDE) describing a truncated Wigner function of a quantum limit-cycling system in a free-carrier cavity.  However, it is not fully consistent with the classical phase reduction theory, because the notions of the asymptotic phase and PSF, which are essential in the classical theory, are not introduced. Consequently, the limit cycle needs to be approximately symmetric for the analysis of synchronization with periodic forcing~\cite{hamerly2015optical}.
Similar phenomenological phase equations, where the phase simply represents the geometric angle of a circular limit cycle, have also been used in several studies on quantum synchronization~\cite{ludwig2013quantum,xu2015conditional, witthaut2017classical, amitai2017synchronization}; however, a systematic phase reduction theory has not been established so far.

In this study, we formulate a general framework of the phase reduction theory for quantum synchronization under the semiclassical approximation. We derive a linearized multi-dimensional semiclassical SDE from a general master equation that describes weakly-perturbed quantum dissipative systems with a single degree of freedom exhibiting stable nonlinear oscillations,
and subsequently reduce it to an approximate one-dimensional classical SDE for the phase variable of the system (see Fig.~\ref{fig_1}).
The derived phase equation has a simple form, characterized by the natural frequency, PSF, and Hessian matrix of the limit cycle in the classical limit, 
and a noise term arising from quantum fluctuations around the limit cycle.
The quantum-mechanical density matrix and power spectrum of the original system can be approximately reconstructed from the reduced phase equation.

On the basis of the reduced phase equation, synchronization dynamics of quantum nonlinear oscillators can be analyzed in detail by using standard techniques for classical nonlinear oscillators~\cite{winfree2001geometry, kuramoto1984chemical, 
	nakao2016phase, 
	ermentrout2010mathematical,ermentrout1996type,brown2004phase}.
As an example, we analyze synchronization of a quantum vdP oscillator under harmonic driving and squeezing. 
In particular, we consider the case with strong squeezing, 
where the oscillation is asymmetric and the analytical solution is not available. 
It is shown that, even in such cases, we can numerically calculate the necessary 
quantities in the classical limit and use them to analyze the synchronization dynamics 
of the original quantum system, provided that the quantum noise and the perturbations given to the oscillator are sufficiently weak.

The rest of this paper is organized as follows; In Sec.~II, the derivation of the approximate phase equation for a quantum limit-cycle oscillator subjected to weak perturbations is given. In Sec.~III, we analyze a quantum vdP oscillator with harmonic driving and squeezing using the derived phase equation. Section ~IV gives concluding remarks, and Appendices provide detailed derivations of the equations and discussions.

%%%%%%%%%%%%%%%%%
%%% Seciton 2 %%%
%%%%%%%%%%%%%%%%%

\section{Theory}

\subsection{Stochastic differential equation for phase-space variables}

We consider quantum dissipative systems with a single degree of freedom interacting with linear and nonlinear reservoirs, 
which has a stable limit-cycle solution in the classical limit and is driven by weak perturbations.
Under the assumption that correlation times of the reservoirs are 
significantly shorter than the time scale of the main system, a Markovian approximation of the reservoirs can be employed and the evolution of the system can be described by a quantum master equation \cite{carmichael2007statistical, gardiner1991quantum},
\begin{align}
\label{eq:me}
\dot{\rho}
= -i[H + \epsilon {\tilde{H}}(t), \rho] 
+ \sum_{m=1}^{n} \mathcal{D}[L_{m}]\rho,
\end{align}
where $\rho$ is a density matrix representing the system state, 
$H$ is a system Hamiltonian,
$\epsilon {\tilde{H}}(t)$ is a time-dependent Hamiltonian 
representing weak external perturbations applied to the system 
($0 < \epsilon \ll 1$), $n$ is the number of reservoirs, $L_{m}$ 
is the coupling operator between the system and 
the $m$th reservoir $(m=1,\ldots,n)$, and $\mathcal{D}[L]\rho = L \rho L^{\dag} - (\rho L^{\dag} L - L^{\dag} L \rho)/2$ denotes the Lindblad form.
We consider a physical condition where the effects of the quantum noise and external perturbations are sufficiently weak and of the same order, and 
perturbatively analyze their effect on the semiclassical dynamics of the system.

First, we transform Eq.~(\ref{eq:me}) into a multi-dimensional SDE by 
introducing a phase-space quasiprobability distribution, 
such as the P, Q, or Wigner representation~\cite{carmichael2007statistical, gardiner1991quantum}. 
In this paper, we use the P representation, because the density matrix and spectrum can be reconstructed using a simple and natural approximation.
In the P representation, the density matrix $\rho$ is represented as
$\rho = \int P({\bm \alpha}) | \alpha \rangle \langle \alpha |  d {\bm \alpha}$,
where $| \alpha \rangle$ is a coherent state specified by 
a complex value $\alpha \in \mathbb{C}$, or equivalently by a two-dimensional complex vector $\bm{\alpha} = (\alpha, \alpha^{*})^{T} \in \mathbb{C}^{2 \times 1}$, $P({\bm \alpha})$ is a quasiprobability 
distribution of ${\bm \alpha}$, $d{\bm \alpha} = d\alpha d\alpha^*$, the integral is taken over the entire space spanned by ${\bm \alpha}$,
and * indicates complex conjugate.

The Fokker-Planck equation (FPE) equivalent to Eq.~(\ref{eq:me}) can be written as
\begin{align}
\label{eq:fpe}
\frac{\pa P(\bm{\alpha}, t)}{\pa t} = \Big[ - \sum_{j=1}^{2} \partial_{j} \{ A_{j}(\bm{\alpha}) + \epsilon \tilde{A}_j(\bm{\alpha}, t) \}
+ \frac{1}{2} \sum_{j=1}^2 \sum_{k=1}^2 \partial_{j}\partial_{k} \{ \epsilon D_{jk}(\bm{\alpha}) \} \Big]P(\bm{\alpha}, t), 
\end{align}
where
$A_j({\bm \alpha})$ and $\tilde{A}_j({\bm \alpha}, t)$ are the $j$th components of complex vectors 
$\bm{A}(\bm{\alpha}) = (A_{1}(\bm{\alpha})$, $A^*_{1}(\bm{\alpha}))^T \in {\mathbb C}^{2 \times 1}$ and
$\bm{\tilde{A}}(\bm{\alpha}, t) = (\tilde{A}_{1}(\bm{\alpha}, t), \tilde{A}^*_{1}(\bm{\alpha}, t))^T \in {\mathbb C}^{2 \times 1}$
representing the system dynamics and perturbations, respectively, 
$ \epsilon D_{jk}({\bm \alpha})$ is the $(j, k)$-th component of the symmetric diffusion matrix $ \epsilon {\bm D}({\bm \alpha}) \in {\mathbb C}^{2 \times 2}$ 
representing quantum fluctuations, and the complex partial derivatives are defined as 
$\partial_1 = \partial / \partial \alpha$ and $\partial_2 = \partial / \partial \alpha^*$
(note that $A_2(\bm{\alpha}) = A^*_{1}(\bm{\alpha})$ and $\tilde{A}_{2}(\bm{\alpha}, t) = \tilde{A}^*_{1}(\bm{\alpha}, t)$).

The drift term $\bm{A}(\bm{\alpha})$ consists of terms arising from the system Hamiltonian $H$ and 
the dissipation $\{ L_m \}$, $\epsilon \tilde{\bm{A}}(\bm{\alpha}, t)$ represents the small terms 
arising from the perturbation Hamiltonian $\epsilon {\tilde{H}}(t)$, and the diffusion matrix 
$ \epsilon {\bm D}({\bm \alpha})$ represents the intensity of the small quantum noise, generally arising from all terms of 
$H$, $ \epsilon {\tilde{H}}(t)$, and $\{ L_m \}$.
These terms can be explicitly calculated from the master equation in Eq.~(\ref{eq:me}) by using the standard calculus for phase-space representation when $H$, $ \epsilon {\tilde{H}}(t)$, and $\{ L_m \}$ are given~\cite{carmichael2007statistical, gardiner1991quantum}.
The external perturbation $\epsilon \tilde{\bm A}({\bm \alpha}, t)$ and the diffusion matrix $\epsilon{\bm D}({\bm \alpha})$ are assumed to be of the same order, $\mathcal{O}(\epsilon)$.

By introducing an appropriate complex matrix $ \sqrt{\epsilon} \bm{\beta}(\bm{\alpha}) \in \mathbb{C}^{2 \times 2}$
(see Appendix \ref{ap_1} for the explicit form),
the diffusion matrix $ \epsilon {\bm D}({\bm \alpha})$ can be represented as
$\epsilon \bm{D}(\bm{\alpha}) = \sqrt{\epsilon} \bm{\beta}(\bm{\alpha})  ( \sqrt{\epsilon} \bm{\beta}(\bm{\alpha}) )^{T}$
and the Ito SDE corresponding to Eq.~(\ref{eq:fpe}) for the phase-space variable ${\bm \alpha}(t)$ is given by
\begin{align}
\label{eq:ldv}
d\bm{ \alpha} &=\{ {\bm A}({\bm \alpha}) + \epsilon \tilde{\bm A}({\bm \alpha}, t) \} dt + 
\sqrt{\epsilon} {\bm \beta}({\bm \alpha}) d{\bm W},
\end{align}
where $\bm{W}(t) = (W_1(t), W_2(t))^T \in \mathbb{R}^{2 \times 1}$ represents a vector of independent Wiener processes $W_{i}(t)~(i=1,\ldots,2)$ satisfying $\mean{dW_{i}dW_{j}} = \delta_{ij} dt$.

It should be noted that diffusion matrix of certain quantum systems in the P representation becomes negative definite
for certain ${\bm \alpha}$~\cite{carmichael2007statistical, gardiner1991quantum}. 
For such systems, we need to employ, for example, the positive P representation with two additional nonclassical variables in place of the P representation, as used by Navarrete-Benlloch {\it et al.}~\cite{navarrete2017general}
in the Floquet analysis of quantum oscillations.
In this study, to present the fundamental idea of the semiclassical phase reduction in its simplest form, 
we only consider the case for which the diffusion matrix is always positive
semidefinite along the limit cycle and formulate the  phase reduction theory in the two-dimensional phase space of classical variables.

\subsection{Derivation of the phase equation}

Our aim is to derive an approximate one-dimensional SDE for the phase variable of the system from the SDE in Eq.~(\ref{eq:ldv}) in the $P$ representation.
To this end, we define a real vector ${\bm X} = (x, p)^T = ( \mbox{Re}\ \alpha, \mbox{Im}\ \alpha)^T \in {\mathbb R}^{2 \times 1}$ from 
the complex vector ${\bm \alpha}$.
The real-valued expression of Eq.~(\ref{eq:ldv}) for ${\bm X}(t)$ is then given by an Ito SDE,
\begin{align}
\label{eq:X}
d{\bm{X}} = \{ {\bm{F}}({\bm X}) + \epsilon {\bm q}({\bm X}, t) \} dt + 
\sqrt{\epsilon} {\bm{G}}({\bm X}) d {\bm W},
\end{align}
where $\bm{F}({\bm X}) \in {\mathbb R}^{2 \times 1}$, ${\bm q}({\bm X}, t) \in {\mathbb R}^{2 \times 1}$, 
and ${\bm{G}}({\bm X}) \in {\mathbb R}^{2 \times 2}$ are real-valued equivalent 
representations of the system dynamics ${\bm A}({\bm \alpha}) \in {\mathbb C}^{2 \times 1}$, 
perturbation $\tilde{\bm A}({\bm \alpha}, t) \in {\mathbb C}^{2 \times 1}$, and noise intensity ${\bm \beta}({\bm \alpha}) 
\in {\mathbb C}^{2 \times 2}$ in Eq.~(\ref{eq:ldv}), respectively.

We assume that the system in the classical limit without perturbation and quantum noise, $\dot{{\bm{X}}} = {\bm{F}}({\bm{X}})$, has an exponentially stable limit-cycle solution ${\bm{X}}_{0}(t) = (x_0(t), p_0(t))^T = {\bm{X}}_{0}(t+T)$ with a natural period $T$ and frequency $\omega = 2\pi / T$.
In the same way as the phase reduction for classical limit cycles
\cite{winfree2001geometry, kuramoto1984chemical, 
	nakao2016phase, 
	ermentrout2010mathematical,ermentrout1996type,brown2004phase},
we can introduce an asymptotic phase function $\Phi({\bm{X}}) : {B} \subset {\mathbb R}^2 \to [0, 2\pi)$ such that
$\nabla \Phi({\bm{X}}) \cdot {\bm F}({\bm{X}})  = \omega$
is satisfied for all system states ${\bm X}$ in the basin ${B}$ of the limit cycle in the classical limit, where $\nabla \Phi({\bm X}) \in {\mathbb R}^{2 \times 1}$ is the gradient of $\Phi({\bm X})$.
Using this phase function, we define the phase of a system state ${\bm X} \in B$ as $\phi = \Phi({\bm X})$. It then follows that $\dot{\phi} = \dot{\Phi}({\bm X}) = {\bm F}({\bm X}) \cdot \nabla \Phi({\bm X}) = \omega$, i.e., $\phi$ always increases at a constant frequency $\omega$ with the evolution of ${\bm X}$.
In the following formulation, we represent the system state ${\bm X}$ on the limit cycle as ${\bm X}_0(\phi) = ( x_0(\phi), p_0(\phi) )^T$ as a function of the phase $\phi$ rather than the time $t$.
In this representation, ${\bm X}_0(\phi)$ is a $2\pi$-periodic function of $\phi$, ${\bm X}_0(\phi) = {\bm X}_0(\phi+2\pi)$.
Note that an identity $\Phi({\bm X}_0(\phi)) = \phi$ is satisfied by the definition of $\Phi({\bm X})$.

When the noise and perturbations are sufficiently weak and the deviation of the state ${\bm X}$ from the limit cycle is small, we can approximate ${\bm X}(t)$ by a state ${\bm X}_0(\phi(t))$ on the limit cycle as ${\bm X}(t) \approx {\bm X}_0(\phi(t))$ and derive a SDE for the phase in the lowest order approximation by using the Ito formula as (see Appendix \ref{ap_2} for details)
\begin{align}
\label{eq:dphi}
d\phi = \left\{ \omega + \epsilon \bm{Z}( \phi ) \cdot {\bm q}(\phi, t) + \epsilon g(\phi) \right\} dt
+ \sqrt{\epsilon} \{ \bm{G}(\phi)^T {\bm Z}(\phi) \} \cdot d\bm{W},
\end{align}
where the drift term is correct up to $\mathcal{O}(\epsilon)$ and the noise intensity is correct up to $\mathcal{O}(\sqrt{\epsilon})$.
Here, the inner product between two vectors 
$\bm{a} = (a_0, a_2, \cdots, a_{N-1})^T \in {\mathbb R}^{N \times 1}$
and
$\bm{b} = (b_0, b_2, \cdots, b_{N-1})^T \in {\mathbb R}^{N \times 1}$
is defined as $\bm{a} \cdot \bm{b} = \sum_{i = 0}^{N-1} a_i b_i$.

In the above phase equation, the gradient $\nabla \Phi$ of $\Phi({\bm X})$ at ${\bm X}$ 
is approximately evaluated at ${\bm X}(\phi)$ on the limit cycle and is denoted as
$\bm{Z}(\phi) = \nabla \Phi|_{ {\bm{X} }  = {\bm{X}}_{0}(\phi)} \in {\mathbb R}^{2 \times 1}$.
We call this ${\bm Z}(\phi)$ the \textit{phase sensitivity function} (PSF) of the limit cycle, which characterizes the linear response property of the oscillator phase to given perturbations~\cite{kuramoto1984chemical,nakao2016phase}.
Similarly, the perturbation and noise intensity can also be evaluated approximately at ${\bm X} = {\bm X}_0(\phi)$ on the limit cycle and 
they are denoted as ${\bm q}(\phi, t) = {\bm q}({\bm X}_{0}(\phi), t)$ and $\bm{G}(\phi) = \bm{G}({\bm X}_{0}(\phi))$, respectively.
The additional function $g(\phi)$ in the drift term in Eq.~(\ref{eq:dphi}) arises from the change of the variables and is given by
\begin{align}
g(\phi) = \frac{1}{2} \mbox{Tr} \left\{ {\bm G}(\phi)^T {\bm Y}(\phi) {\bm G}(\phi) \right\},
\end{align}
where ${\bm Y}(\phi) = \nabla^T \nabla \Phi|_{{\bm X} = {{\bm X}_0(\phi) }} $ is a Hessian matrix of the phase function $\Phi({\bm X})$ also evaluated at ${\bm X} = {\bm X}_0(\phi)$ on the limit cycle.
All these functions are $2\pi$-periodic, as they are functions of ${\bm X}_0(\phi)$.

It is well known in the classical phase reduction theory that the PSF can be obtained as a $2 \pi$-periodic solution
to the following adjoint equation and an additional normalization condition~\cite{ermentrout1996type,brown2004phase,nakao2016phase}:
\begin{align}
\label{eq:adjoint}
\omega \frac{d}{d\phi} { \bm{Z}}(\phi)  = - {\bm J}^T( \phi ) 
\bm{Z}(\phi),
\quad
\bm{Z}(\phi) \cdot \frac{d \bm{X}_{0}(\phi)}{d\phi}= 1,
\end{align}
respectively, where ${\bm J}(\phi) = {\bm J}({\bm X}_0(\phi)) \in {\mathbb R}^{2 \times 2}$ is a Jacobian matrix of ${\bm F}({\bm X})$ at ${\bm X} = {\bm X}_0(\phi)$ on the limit cycle.
It is also known that the Hessian matrix ${\bm Y}(\phi)$ on the limit cycle can be calculated as a $2 \pi$-periodic solution of an adjoint-type equation~\cite{suvak2010quadratic, takeshita2010higher} with an appropriate constraint.
These equations for ${\bm Y}(\phi)$ are detailed in the Appendix~\ref{ap_2}.
In the numerical calculations, ${\bm Z}(\phi)$ can easily be obtained by 
the backward integration of the adjoint equation with occasional normalization as proposed by Ermentrout~\cite{ermentrout2010mathematical},
and then the Hessian ${\bm Y}(\phi)$ can be obtained by a shooting method~\cite{suvak2010quadratic}.

Because of the additional term $g(\phi)$ in Eq.~(\ref{eq:fpephi}), the {\it effective} frequency $\tilde{\omega} = \langle d \phi \rangle / dt$ of the oscillator in the absence of the perturbation ${\bm q}(\phi, t)$ is given by
\begin{align}
\tilde{\omega} = \omega + \frac{\epsilon}{2\pi} \int_0^{2\pi} g(\psi') d\psi',
\label{eq:effectivefreq}
\end{align}
which is slightly different from the natural frequency of the oscillator $\omega$ in the classical limit. 
Though not used in the present study, we can further introduce a new phase variable $\psi$ that is only slightly different from $\phi$ by a near-identity transform as $\phi = \psi + \epsilon n(\psi)$, where $n(\psi)$ is a $2\pi$-periodic function with $n(0) = 0$, and eliminate the additional function $g(\phi)$ in Eq.~(\ref{eq:dphi}) by renormalizing it into the frequency term. The new phase $\psi$ then obeys a simpler SDE of the form
\begin{align}
d\psi = \{ \tilde{\omega} + \epsilon \bm{Z}( \psi ) \cdot {\bm q}(\psi, t) \} dt
+ \sqrt{\epsilon} h(\psi) dW,
\label{eq:averagedphase}
\end{align}
where $h(\psi) = \sqrt{ \sum_{i=1}^2 \left\{ {\bm G}(\psi)^T {\bm Z}(\psi) \right\}_i^2 }$ and $W(t)$ is a one-dimensional Wiener process. As before, the drift term is correct up to $\mathcal{O}(\epsilon)$ and the noise intensity is correct up to $\mathcal{O}(\sqrt{\epsilon})$.
See Appendix \ref{ap_3} for the details.
In this study, we use the original phase equation in Eq.~(\ref{eq:dphi}) for numerical simulations and verify its validity.
We also note here that the phase equation derived in Ref.~\cite{hamerly2015optical} does not contain a term with the Hessian matrix, because the order of the noise intensity is implicitly assumed to be $\mathcal{O}(\epsilon)$ in~\cite{hamerly2015optical}.

From the reduced SDE in Eq.~(\ref{eq:dphi}), we can derive a corresponding FPE describing the probability density function 
$P(\phi, t)$ of the phase variable $\phi$ as
\begin{align}
\frac{\partial}{\partial t} P(\phi, t) =
- \frac{\partial}{\partial \phi} \left\{ \omega + \epsilon {\bm Z}( \phi ) \cdot {\bm q}( \phi, t ) + \epsilon g(\phi) \right\} P(\phi, t)
+ \frac{\epsilon}{2} \frac{\partial^2}{\partial \phi^2} h(\phi)^2 P(\phi, t).
\label{eq:fpephi}
\end{align}
Using this FPE, we can obtain the stationary distribution and transition probability of the phase variable $\phi$ and use them to reconstruct the density matrix and power spectrum.

\subsection{Reconstruction of the density matrix}

From the reduced phase equation, we can approximately reconstruct the quantum state as follows.
Using the phase variable $\phi$, the oscillator state in the classical limit can be approximated as ${\bm X} \approx {\bm X}_0(\phi) = (x_0(\phi), p_0(\phi))^T$, 
or
${\bm \alpha} \approx {\bm \alpha}_0(\phi) = (\alpha_0(\phi), \alpha_0(\phi)^*)^T
= (x_0(\phi) + i p(\phi), x_0(\phi) - i p(\phi) )^T$
in the original complex representation.
Therefore, the quantum state at phase $\phi$ is approximately described as $\left| \alpha_0(\phi) \right\rangle$
and the density matrix $\rho$ is approximately represented by
using the probability density function $P(\phi)$ of the phase variable $\phi$,
obtained from the SDE in Eq.~(\ref{eq:dphi}) or FPE in Eq.~(\ref{eq:fpephi}), as
\begin{align}
\label{eq:rho}
\rho \approx \int_{0}^{2 \pi}d \phi P(\phi)
\left| \alpha_{0}(\phi) \right\rangle \left \langle \alpha_{0}(\phi) \right|,
\end{align}
which is simply a mixture of coherent states weighted by the distribution of the phase on the classical limit cycle.
Thus, we can approximately reconstruct the density matrix of the original quantum oscillator 
from the classical SDE for the phase variable $\phi$,
which is characterized by the natural frequency $\omega$, PSF ${\bm Z}(\phi)$, Hessian matrix ${\bm Y}(\phi)$, and noise intensity ${\bm G}(\phi)$ that represents quantum fluctuations around the limit cycle.

The derivation of the phase equation in Eq.~(\ref{eq:dphi}) from the original quantum-mechanical master equation in Eq.~(\ref{eq:me}) and reconstruction of the quantum-mechanical density matrix from the approximate phase equation, Eq.~(\ref{eq:rho}), are the main result of the present work. 
A schematic diagram of the proposed method is illustrated in Fig.~\ref{fig_1}.
The reduced phase equation is essentially the same as that for the classical limit-cycle oscillator driven by noise, and synchronization dynamics of the weakly perturbed quantum nonlinear oscillator in the semiclassical regime can be analyzed on the basis of the reduced phase equation by using the standard methods for the classical limit-cycle oscillator.

%%%%%%%%%%%%%%%%%
%%% Seciton 3 %%%
%%%%%%%%%%%%%%%%%

\section{Examples}

\subsection{Quantum van der Pol oscillator with harmonic driving and squeezing}

As an example, we consider a quantum vdP oscillator subjected to harmonic driving and squeezing.
We assume that the harmonic driving is sufficiently weak and treat it as a perturbation.
As for the squeezing, we consider two cases; 
(i) the squeezing is sufficiently weak and can also be treated as a perturbation, and
(ii) the squeezing is relatively strong and cannot be treated as a perturbation.

We denote by $\omega_{0}$, $\omega_{d}$, and $\omega_{sq}$ the frequencies of the oscillator, harmonic driving, and pump beam of squeezing, respectively. 
We consider the case where the squeezing is generated by a degenerate parametric amplifier and assume $\omega_{sq} = 2\omega_{d}$ \cite{gardiner1991quantum}.
In the rotating coordinate frame of frequency $\omega_{d}$, the master equation
is given by~\cite{walter2014quantum,sonar2018squeezing}
\begin{align}
\label{eq:qvdp_me}
\dot{\rho} = 
-i \left[  - \Delta a^{\dag}a + i E (a - a^{\dag}) 
+ i \eta ( a^2 e^{-i \theta} - a^{\dag 2} e^{ i \theta}  )
, \rho \right]
+ \gamma_{1} \mathcal{D}[a^{\dag}]\rho + \gamma_{2}\mathcal{D}[a^{2}]\rho,
\end{align}
where $\Delta = \omega_{d} - \omega_{0}$ is the frequency detuning of
the harmonic driving from the oscillator,  $E$ is the intensity of the
harmonic driving, $\eta e^{ i \theta}$ is the squeezing parameter, 
$\gamma_{1}$ and $\gamma_{2}$ are the decay rates for 
negative damping and nonlinear damping, respectively,
and the Planck constant is set as $\hbar = 1$.
The harmonic driving is represented by a constant $E$, because a coordinate frame rotating with the driving frequency $\omega_d$ is used.

We assume that $\gamma_{2}$ is sufficiently small and of $\mathcal{O}(\epsilon)$, for which the semiclassical approximation is valid, and represent $\gamma_2$ as $\gamma_{2} = \epsilon \gamma_{1} \gamma_{2}{'}$
using a dimensionless parameter $\gamma_{2}{'}$ of $\mathcal{O}(1)$.
In this setting, the size of the stable limit-cycle solution in Eq.~(\ref{eq:qvdp_me}) 
in the classical limit is  $\mathcal{O}(1 / \sqrt{\epsilon})$, 
while we have implicitly assumed it to be $\mathcal{O}(1)$ in the derivation of Eq.~(\ref{eq:dphi}).
Therefore, we introduce a rescaled annihilation operator $a'$ and the corresponding classical variable $\alpha'$ (${\bm \alpha}' = (\alpha', \alpha'^{*})$ in the vector representation) as $ a' | \alpha' \rangle = \sqrt{\epsilon} a| \sqrt{\epsilon} \alpha \rangle$, and represent the parameters as  $\Delta = \gamma_{1} \Delta{'}, E = \sqrt{\epsilon} \gamma_{1} E{'}, \eta = \delta \gamma_{1} \eta{'}$, where $\Delta', E'$, and $\eta'$ are dimensionless parameters of $\mathcal{O}(1)$.
By this rescaling, the size of the limit cycle becomes $\mathcal{O}(1)$ and
the parameter $\delta$ determines the relative intensity of the squeezing.

The real-valued representation 
$\bm{X} = (x', p')^T = (\mbox{Re}~\alpha', \mbox{Im}~\alpha')^T$
of Eq.~(\ref{eq:X}) after rescaling is then obtained as
\begin{align}
\label{eq:drift}
d \bm{X}
=
\begin{pmatrix}
\frac{1}{2}x'  - \Delta' p'  
- \gamma'_{2} x' (x'^{2} + p'^{2}) 
-\epsilon E' - 2\delta \eta' ( x' \cos \theta + p' \sin \theta)  
\cr
\frac{1}{2}p' + \Delta' x'  
- \gamma'_{2} p' (x'^{2} + p'^{2}) 
+ 2\delta \eta' ( p' \cos \theta - x' \sin \theta) 
\end{pmatrix}
dt'
+\sqrt{\epsilon} \bm{G}(\bm{X})
d\bm{W}',
\end{align}
where $dt' = \gamma_1dt$ and $d\bm{W}' = \sqrt{\gamma_{1}}d\bm{W}$. The noise intensity matrix is explicitly given by 
\begin{align}
\label{eq:noiseintensity}
\bm{G}(\bm{X}) & =
\left( \begin{matrix}
\sqrt{\frac{\left( 1 +  R'_{1} \right)}{2}} 
\cos \frac{\chi'_1}{2} & 
\sqrt{\frac{\left( 1 -  R'_{1} \right)}{2}} 
\sin \frac{\chi'_1}{2} \\
\sqrt{\frac{\left( 1 +  R'_{1} \right)}{2}} 
\sin \frac{\chi'_1}{2} & 
- \sqrt{\frac{\left( 1 -  R'_{1} \right)}{2}}  
\cos \frac{\chi'_1}{2} \\
\end{matrix} \right),
\end{align}
with $R'_{1}e^{i \chi'_1} = -(\gamma'_2 (x' + ip')^2 + 2 \delta \eta' e^{i\theta})$.
Further details of the derivation can be found in the Appendix \ref{ap_5}.

\subsection{Weak squeezing}

First, we consider the case of weak squeezing with $\delta=\epsilon$.
The rescaled system and perturbation Hamiltonians are given by
\begin{align}
H = - \Delta' a'^{\dag}a',
\quad
\epsilon {\tilde{H}} = \epsilon \{ i  E' (a' - a'^{\dag})
+ i \eta' ( a'^2 e^{-i \theta} - a'^{\dag 2} e^{ i \theta}) \}.
\end{align}
For this system, we obtain ${\bm F}({\bm X}) = ( {x}'/2  - \Delta' {p}'  - \gamma'_{2} {x}' ({x}'^{2} + {p}'^{2}),\ {p}'/2  + \Delta' {x}'  - \gamma'_{2} {p}' ({x}'^{2} + {p}'^{2}) )^T$.
The perturbation is represented by ${\bm q}({\bm X}, t) = ( -E' - 2 \eta' ( x' \cos \theta + p' \sin \theta),\ 2 \eta' ( p' \cos \theta - x' \sin \theta)  )^T$.
Note that the vector field ${\bm F}({\bm X})$ in this case is simply a normal form of the supercritical Hopf bifurcation. A classical nonlinear oscillator described by this ${\bm F}({\bm X})$ is known as the {\it Stuart-Landau} (SL) oscillator~\cite{kuramoto1984chemical} (which is different from the classical vdP oscillator) and it is analytically solvable.

The stable limit cycle of the SL oscillator is given by
\begin{align}
{\bm X}_0(\phi) = \sqrt{\frac{1}{2\gamma'_{2}}} \begin{pmatrix} \cos \phi \\ \sin \phi \end{pmatrix}
\end{align}
as a function of phase $\phi = \omega t$, where the frequency is given by $\omega = \Delta'$.
The basin $B$ of this limit cycle is the whole $(x', p')$-plane except $(0, 0)$.
The phase function $\Phi(\bm{X})$ of this limit cycle can be expressed as $\Phi(x', p') = \tan^{-1} (p' / x') $~\cite{nakao2016phase}, which gives $\dot{\phi} = \dot{\Phi}(x', p') = \omega$.
The PSF ${\bm Z}(\phi)$ and Hessian matrix ${\bm Y}(\phi)$ can be obtained by calculating the gradients of the phase function $\Phi({\bm X})$ at ${\bm X} = {\bm X}_0(\phi)$ on the limit cycle as
\begin{align}
{\bm Z}(\phi) = \sqrt{2\gamma'_{2}} \begin{pmatrix} -\sin \phi \\ \cos \phi \end{pmatrix},
\quad
{\bm Y}(\phi) = 2 \gamma'_{2} \begin{pmatrix} \sin 2\phi & - \cos 2\phi \\ - \cos 2\phi & - \sin 2\phi \end{pmatrix}.
\end{align}
In this case, the additional term $g(\phi)$ in Eq.~(\ref{eq:dphi}) and therefore the frequency shift in Eq.~(\ref{eq:effectivefreq}) vanishes, i.e., $\tilde{\omega} = \omega$.
The $\mathcal{O}(\epsilon \sqrt{\epsilon})$ terms in the noise intensity ${\bm G}(\phi)$ given by Eq.~(\ref{eq:noiseintensity}) are neglected.

From these results, the phase equation in Eq.~(\ref{eq:dphi}) for the quantum vdP oscillator driven by weak harmonic driving and squeezing is explicitly given by
\begin{align}
\label{eq:qvdp_phi}
d\phi &= \left\{ \Delta' +  \sqrt{2} \epsilon \sqrt{\gamma'_{2}}E'\sin \phi + 
2 \epsilon \eta' \sin (2 \phi - \theta) \right\} dt'
+ \sqrt{\epsilon} \sqrt{\frac{3 \gamma'_{2}}{2}} dW'
\end{align}
in the lowest-order approximation, where $dW'= \sqrt{\gamma_{1}}dW$.
Using the probability density function $P(\phi)$ of the phase $\phi$ described by the FPE~(\ref{eq:fpephi}) corresponding to Eq.~(\ref{eq:qvdp_phi}), the approximate density matrix, Eq.~(\ref{eq:rho}), is explicitly given by
\begin{align}
\rho \approx \int_{0}^{2 \pi}d \phi P(\phi)
\left|
\sqrt{ \frac{\gamma_{1}}{2\gamma_{2}}} e^{ i \phi}
\right\rangle 
\left\langle 
\sqrt{ \frac{\gamma_{1}}{2\gamma_{2}}} e^{ i \phi}
\right|.
\end{align}

\subsection{Strong squeezing}

Next, we consider the case of strong squeezing with $\delta = 1$ and incorporate it into the system Hamiltonian.
The rescaled system and perturbation Hamiltonians are given by
\begin{align}
H = - \Delta' a'^{\dag}a' + i \eta' ( a'^2 e^{-i \theta} - a'^{\dag 2} e^{ i \theta}),
\quad
\epsilon {\tilde{H}} = \epsilon i  E' (a' - a'^{\dag}).
\end{align}
We obtain ${\bm F}({\bm X}) = ( {x}'/2  - \Delta' {p}'  - \gamma'_{2} {x}' ({x}'^{2} + {p}'^{2})- 2 \eta' ( {x}' \cos \theta + {p}' \sin \theta),\; {p}'/2  + \Delta' {x}'  - \gamma'_{2} {p}' ({x}'^{2} + {p}'^{2}) + 2 \eta' ({p}' \cos \theta - {x}' \sin \theta))^T$ with extra terms due to squeezing, characterized by the parameter $\eta'$.
When $\Delta' > 2 \eta'$ (i.e., $\Delta > \ 2 \eta$), this vector field ${\bm F}({\bm X})$ possesses a stable limit-cycle solution ${\bm X}_0(t)$ in the classical limit.
Due to the strong squeezing, this limit cycle is asymmetric and the angular velocity of the oscillator state is non-uniform. At $\Delta' = 2 \eta'$, this limit cycle disappears via a saddle-node bifurcation on invariant circle. 
The perturbation is given by ${\bm q}({\bm X}, t) = ( -E', 0 )$.

In this case, the system is not analytically solvable, but we can numerically obtain the limit-cycle solution ${\bm X}_0(\phi) = (x_0(\phi), p_0(\phi))^T$, natural frequency $\omega$, PSF $\bm Z(\phi)$, and Hessian matrix ${\bm Y}(\phi)$, and use them in the phase equation in Eq.~(\ref{eq:dphi}).
The density matrix can be approximately reconstructed from Eq.~(\ref{eq:rho}), where $\alpha_0(\phi) = x_0(\phi) + i p_0(\phi)$.
In this case, the frequency shift does not vanish generally and the effective frequency $\tilde{\omega}$ is slightly different from $\omega$ in the classical limit without noise.

An example of the limit cycle in the classical limit is shown in Fig.~\ref{fig_3}(c), and the PSF is shown in Fig.~\ref{fig_3}(d) and (e).
The effective frequency is evaluated as $\tilde{\omega} = 0.7743$  at the parameter values given in Fig.~\ref{fig_3}, which is slightly different from the natural frequency $\omega = 0.7746$ of the system in the classical limit without noise.
From the phase equation, we can obtain the stationary phase distribution $P(\phi)$ by solving the corresponding FPE and reconstruct the density matrix as a mixture of the coherent states on the limit cycle.

\subsection{Reconstruction of density matrices}

To test the validity of the reduced phase equation, we compare
the density matrix $\rho_{sc}$,
which is reconstructed from Eq.~(\ref{eq:rho}) by using $P(\phi)$ obtained from the FPE in Eq.~(\ref{eq:fpephi}) associated with the reduced phase equation
in Eq.~(\ref{eq:dphi}),
with the true density matrix $\rho_{qm}$,
which is obtained by direct numerical simulation of the original quantum master equation in Eq.~(\ref{eq:qvdp_me}),
in the steady state of the system.
We use the fidelity $F= \Tr[ \sqrt{ \sqrt{\rho_{sc}} \rho_{qm}  \sqrt{\rho_{sc}} } ]$~\cite{nielsen2000quantum}
to quantify the similarity between $\rho_{sc}$ and $\rho_{qm}$.
Numerical simulations of the master equation have been performed by using QuTiP~\cite{johansson2012qutip,*johansson2013qutip}
numerical toolbox.

\begin{figure} [!t]
	\begin{center}
		\includegraphics[width=0.6\hsize,clip]{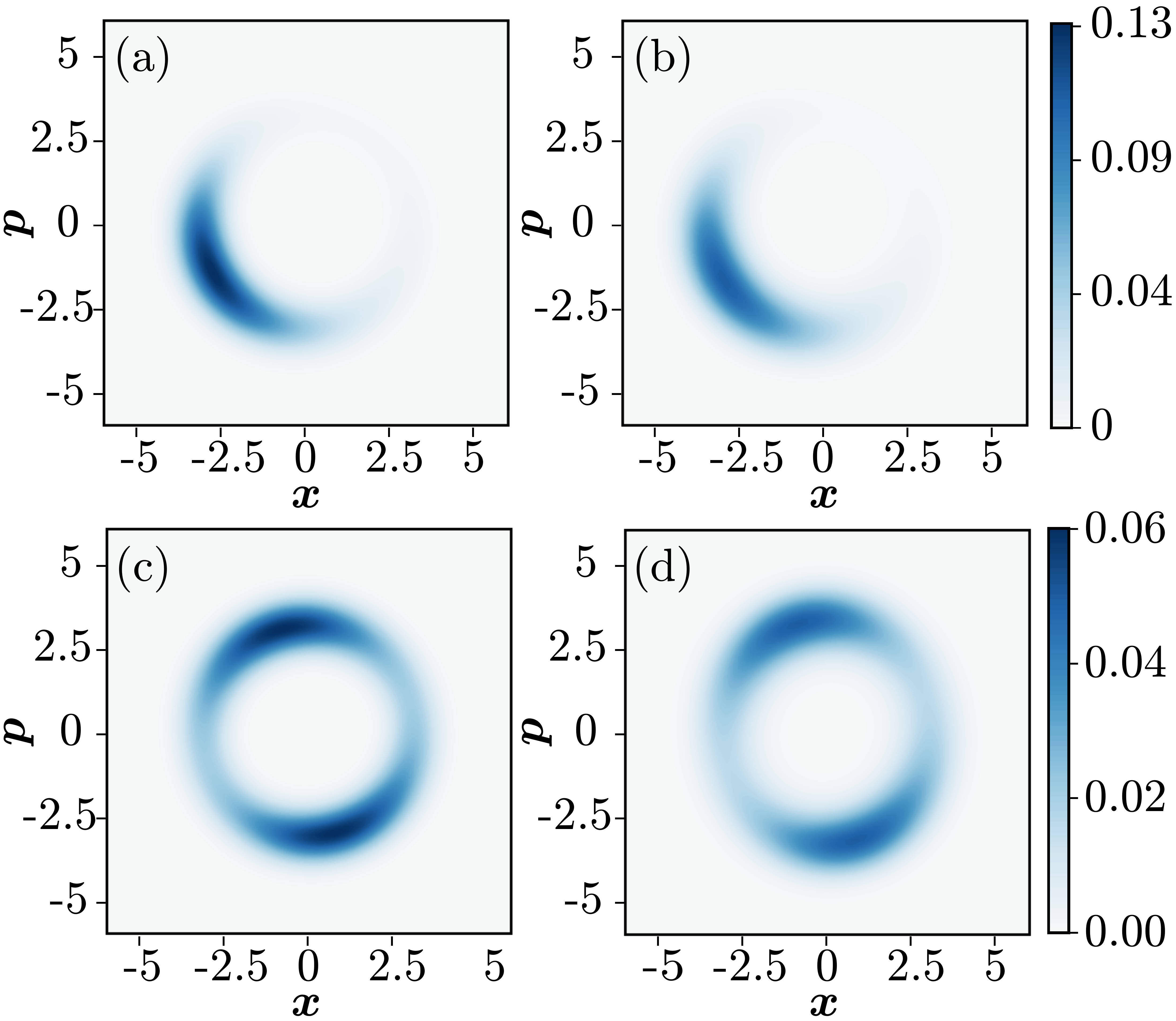}
		\caption{
			Results for the quantum van der Pol oscillator under harmonic driving (a, b) and under weak squeezing (c,d).
			(a,c): Wigner distributions of $\rho_{sc}$ reconstructed from the reduced phase equation, and
			(b,d): Wigner distributions of $\rho_{qm}$ obtained by direct numerical simulation of the original master equation.
			In (a,b), weak harmonic driving with $(\Delta, \gamma_{2}, \eta e^{i \theta}, E)/\gamma_{1} = (0.05, 0.05, 0, \sqrt{0.1})$ is applied,
			and in (c,d), weak squeezing with $(\Delta, \gamma_{2}, \eta e^{i \theta}, E)/\gamma_{1} = (0.05, 0.05, 0.025, 0)$ is applied.
			The fidelities between $\rho_{sc}$ and $\rho_{qm}$ are $F=0.963$ in (a,b) and $F=0.982$ in (c,d), respectively. 
			Note that the figures are drawn using $x$ and $p$ before rescaling.}
		\label{fig_2}
	\end{center}
\end{figure}

Figure~\ref{fig_2}(a)-(d) show the steady-state Wigner distributions corresponding to 
$\rho_{sc}$ and $\rho_{qm}$ under the weak harmonic driving or the squeezing.
In both cases, the distribution is localized along the limit cycle in the classical limit, where the width of the distribution is determined by the intensity of the quantum noise.
In Fig.~\ref{fig_2}(a) and (b), only the harmonic driving is given as the perturbation $(\eta = 0)$, while in Fig.~\ref{fig_2}(c) and (d), only the squeezing is given as the perturbation ($E = 0$).
It can be seen that the true density matrix $\rho_{qm}$ is accurately approximated by the density matrix $\rho_{sc}$ reconstructed from the phase equation in both cases.
The fidelity is $F=0.963$ in the former case and $F=0.982$ in the latter case.

It is notable that the Wigner distribution is localized around one phase point on the limit cycle in Fig.~\ref{fig_2}(a) and (b), which indicates that 
there is a 1:1 phase locking \cite{pikovsky2001synchronization} between the oscillator and the harmonic driving;
In the classical limit, the phase is locked to the point where the deterministic part of Eq.~(\ref{eq:qvdp_phi}) vanishes, and thus the Wigner distribution takes large values around such a point.
Similarly, the Wigner distribution is localized around two phase points on the cycle shown in Fig.~\ref{fig_2}(c) and (d), 
because the frequency of the squeezing is twice that of the harmonic driving and 1:2 phase locking occurs, as can 
be expected from the third term in the deterministic part of Eq.~(\ref{eq:qvdp_phi}) representing the effect of the squeezing. 
Note that Fig.~\ref{fig_2} is depicted in the rotating coordinate frame of frequency $\omega_{d}$ and
the locked phase rotates with frequency $\omega_{d}$ in the original coordinate.

\begin{figure} [!t]
	\begin{center}
		\includegraphics[width=0.75\hsize,clip]{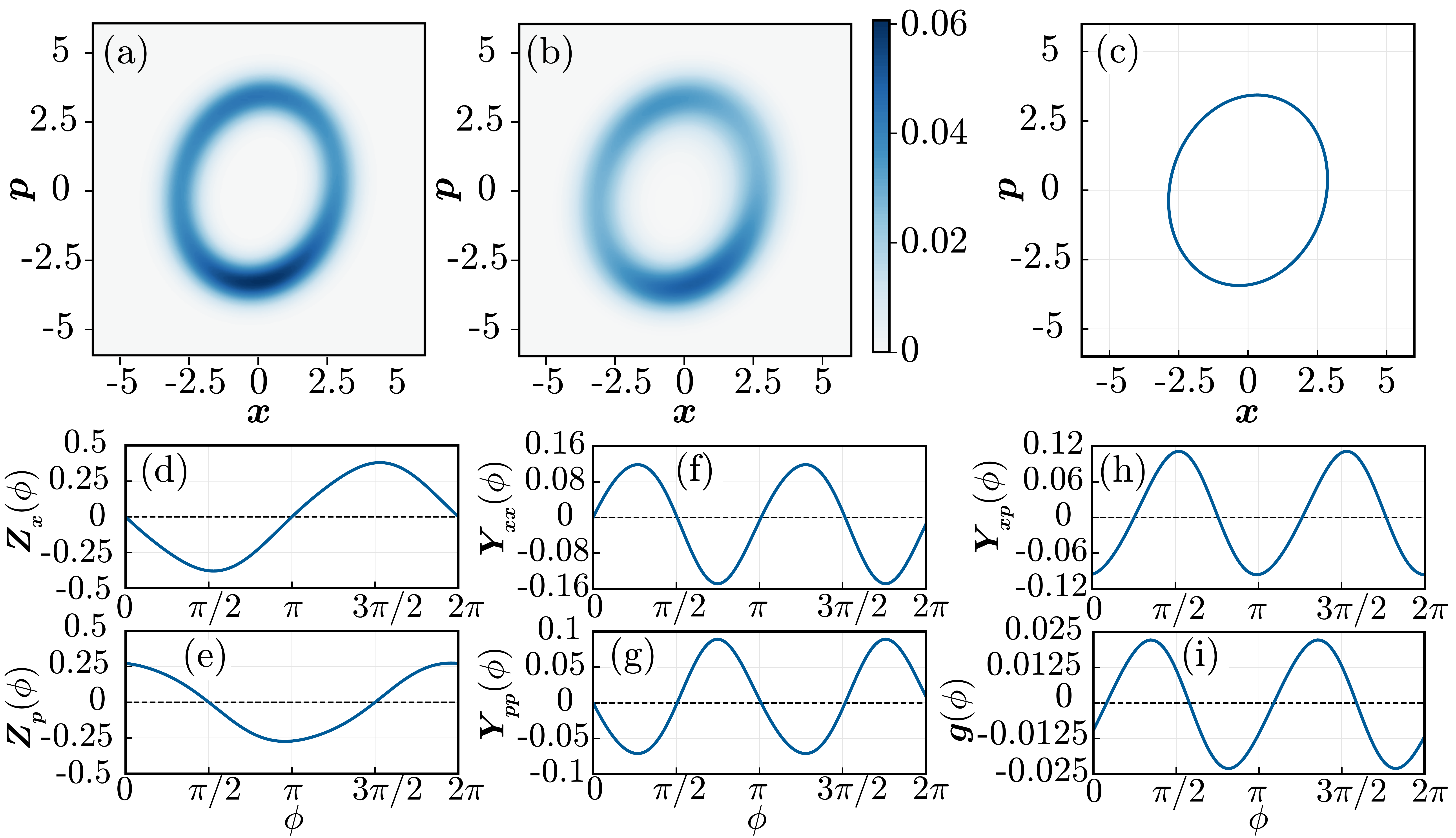}
		\caption{
			Results for the quantum van der Pol oscillator under 
			strong squeezing and weak harmonic driving with parameters $(\Delta, \gamma_{2}, \eta e^{i \theta}, E)/\gamma_{1} = 
			(0.8, 0.05, -0.1i, \sqrt{0.1})$.
			(a): Wigner distribution of $\rho_{sc}$ reconstructed from the reduced phase equation.
			(b): Wigner distribution of $\rho_{qm}$ obtained by direct numerical simulation of the original master equation.
			(c): Limit cycle ${\bm X}_0(\phi) = ( x_0(\phi), p_0(\phi) )^T$ in the classical limit. 
			(d, e): The $x = \mbox{Re}~\alpha$ and $p = \mbox{Im}~\alpha$ components of the PSF ${\bm Z}(\phi)$.
			(f, g, h): The $xx$, $pp$, $xp$ component of the Hessian matrix ${\bm Y}(\phi)$.
			(i): Additional term $g(\phi)$ arising form the change of variables.
			In (a,b), the fidelity between $\rho_{sc}$ and $\rho_{qm}$ is $F=0.976$. }
		\label{fig_3}
	\end{center}
\end{figure}

Figure~\ref{fig_3}(a) and (b) show the Wigner distributions in the case of strong squeezing and weak harmonic driving,  where all quantities are calculated numerically.
In this case, the system exhibits a stable limit cycle in the rotating coordinate frame of
frequency $\omega_{d}$, and constant driving is applied on the the system as in Eq.~(\ref{eq:drift}).
The limit cycle in the classical limit is shown in Fig.~\ref{fig_3}(c),
the $x$ and  $p$ components of the PSF obtained from Eq.~(\ref{eq:adjoint}) are shown in Figs.~\ref{fig_3}(d) and (e), the $xx$, $pp$, $xp$ components of the Hessian matrix are shown in Fig.~\ref{fig_3}(f),(g), and (h) (the $px$ component is equal to the $xp$ component), and the additional term $g(\phi)$ is shown in Fig.~\ref{fig_3}(i).
The origin of the phase $\phi=0$ is chosen as the intersection of the limit cycle and 
the ${x}'$ axis with ${x}'>0$.

It can be seen that the limit cycle in the classical limit is asymmetric due to the effect of the strong squeezing.
The density matrix $\rho_{sc}$ can be reconstructed from the phase distribution $P(\phi)$ obtained numerically.
As shown in Fig.~\ref{fig_3}(a) and (b), the true density matrix $\rho_{qm}$ is well approximated by $\rho_{sc}$ with fidelity $F=0.976$.
In Fig.~\ref{fig_3}(a) and (b), the Wigner distribution is concentrated around the
stable phase point where the deterministic part of the phase equation vanishes.
Thus, the reduced phase equation well reproduces the density matrix of the original quantum system also in this case.

%%%%%%%%%%%%%%%%%
%%% Seciton 5 %%%
%%%%%%%%%%%%%%%%%

\begin{figure} [!t]
	\begin{center}
		\includegraphics[width=0.75\hsize,clip]{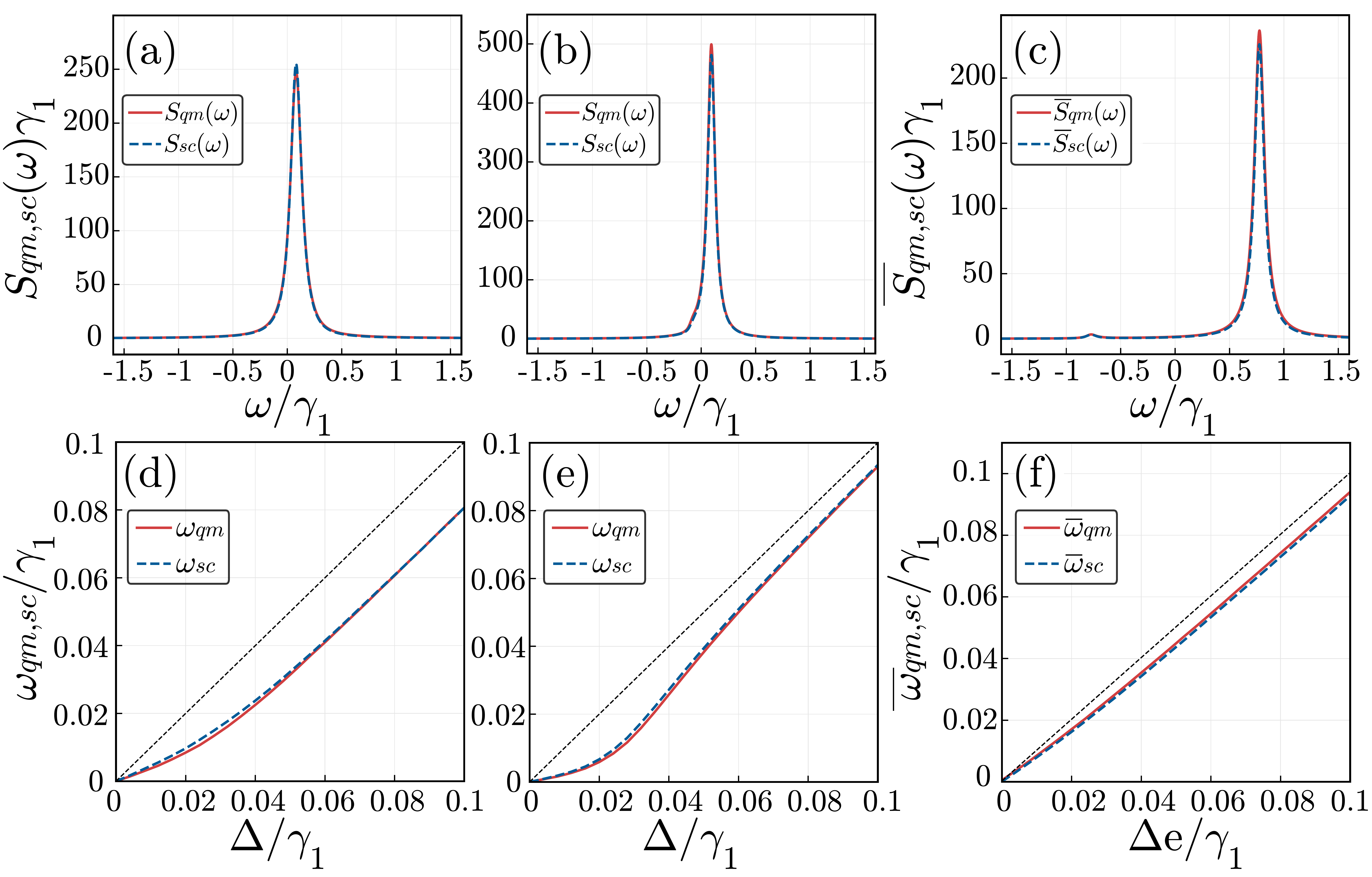}
		\caption{
			Power spectra (a-c) and observed frequencies (d-f) obtained by direct numerical simulations of the master equation (red solid lines) and
			obtained from the reduced phase equation (blue dotted lines).
			(a,d): Weak harmonic driving without squeezing, $(\gamma_{2}, \eta e^{i \theta}, E)/\gamma_{1} = (0.05, 0, \sqrt{0.1})$. $\Delta = 0.1$ in (a).
			(b,e): Weak squeezing without harmonic driving, $(\gamma_{2}, \eta e^{i \theta}, E)/\gamma_{1} = (0.05, 0.025, 0)$. $\Delta = 0.1$ in (b).
			(c,f): Strong squeezing and weak harmonic driving, $(\Delta, \gamma_{2}, \eta e^{i \theta}, E)/\gamma_{1} = 
			(0.8, 0.05, -0.1i, \sqrt{0.1})$. $\Delta_{e} = 0.1$ in (c). In (d-f), the black-dotted lines 
			correspond to the unperturbed cases.
		}
		\label{fig_4}
	\end{center}
\end{figure}

\subsection{Reconstruction of spectra and observed frequencies}

The power spectrum $S_{qm}$ of the original quantum system in the steady state is defined as
\begin{align}
\label{eq_spec_qm}
S_{qm}(\omega)
&= \int_{-\infty}^{\infty} d\tau e^{i\omega \tau} R_{qm}(\tau),
\cr
R_{qm}(\tau) &=
\mean{ a^{\dag}( \tau ) a(0)}_{qm} - \mean{a^{\dag}( \tau )}_{qm} \mean{a(0)}_{qm},
\end{align}
where $R_{qm}$ is the autocovariance and 
$\mean{ A}_{qm} = \Tr{[ A\rho_{qm}]}$ represents the expectation value of an operator $A$ with respect to the steady state density matrix $\rho_{qm}$ obtained from the master equation in Eq.~(\ref{eq:qvdp_me}). 
From the reduced phase equation, using the correspondence between the operators 
and c-numbers in the P representation,
the power spectrum in Eq.~(\ref{eq_spec_qm}) under the semiclassical approximation 
can be reconstructed as
\begin{align}
S_{sc}(\omega)
&= \int_{-\infty}^{\infty} d\tau e^{i\omega \tau} R_{sc}(\tau),
\cr
R_{sc}(\tau) &=
\mean{ \alpha_{0}^{*}( \phi_{2}(\tau) ) \alpha_{0}( \phi_{1}(0) )}_{sc} 
- \mean{\alpha_{0}^{*}( \phi_{2}(\tau) )}_{sc} \mean{\alpha_{0}
	( \phi_{1}(0) )}_{sc}.
\end{align}
Here, $R_{sc}$ is the autocovariance reconstructed from the phase equation,
the mean of a $2\pi$-periodic function $B(\phi)$ is given by
$\mean{ B(\phi) }_{sc} = \int_{0}^{2 \pi}d \phi B(\phi) P_{sc}(\phi)$,
and the autocorrelation is given by 
$\mean{B( \phi_{2}(\tau) ) B( \phi_{1}(0) )}_{sc}$ $= \int_{0}^{2 \pi}d \phi_{1} \int_{0}^{2 \pi}d \phi_{2}( B (\phi_{2}(\tau)) B (\phi_{1}(0)))$ $P(\phi_2, \tau | \phi_1, 0) P_{sc}(\phi_1)$, where $P_{sc}(\phi)$ is a steady phase distribution and $P(\phi_2, t_2 | \phi_1, t_1)$ is a transition probability.
Both of these probability distributions can be calculated from Eq.~(\ref{eq:fpephi}).
The observed frequency $\omega_{qm}$ of the original system and its approximation $\omega_{sc}$ by the phase reduction can be evaluated from the maxima of the spectra as
$\omega_{qm,sc} = \argmax_{\omega} S_{qm,sc}(\omega)$, respectively.

First, we consider the cases with weak squeezing.
Figure~\ref{fig_4}(a) shows the two power spectra $S_{qm}$ and $S_{sc}$ for the case
where only the harmonic driving is given, and Fig.~\ref{fig_4}(b) shows the spectra for
the case with squeezing only.
In both cases, the true spectrum $S_{qm}$ can be accurately approximated by the reconstructed spectrum $S_{sc}$.
The dependence of the observed frequencies $\omega_{qm, sc}$ on the parameter $\Delta$, where $\Delta$ determines the natural
frequency of the limit cycle in the classical limit, is shown in Fig.~\ref{fig_4}(d) and (e).
It can be confirmed that $\omega_{qm}$ is accurately approximated by $\omega_{sc}$ in both cases.
The oscillator strictly synchronizes to the external driving
when the frequency of the oscillator vanishes in the classical limit,
because the harmonic driving acts as a constant force in the rotating frame.
Here, strict synchronization is prevented by the quantum noise and the observed frequencies $\omega_{qm,sc}$ do not vanish completely; however, the tendency toward synchronization can be clearly seen from the decrease in the observed frequency
compared to that of the unperturbed case.

Next, we consider the case with strong squeezing, where
the system exhibits asymmetric limit cycle in the classical limit 
when $\Delta > 2\eta$.
We cannot analyze synchronization with the harmonic driving as a stationary problem by using a rotating coordinate frame of frequency $\omega_{d}$, because the limit cycle is asymmetric and the variation in $\Delta$ does not correspond directly to the variation in $\omega_{d}$.
We thus explicitly apply harmonic driving with periodic amplitude modulation $  E \cos \omega_{e} t$ of frequency $\omega_e$ and measure $\omega_{qm}$ and $\omega_{sc}$ as functions of $\Delta_e = \omega - \omega_e$ for $0 \leq \Delta_e \leq 0.1~ (\omega - 0.1 \leq \omega_e \leq \omega)$, where $\omega = 0.7746$.

\begin{figure} [!t]
	\begin{center}
		\includegraphics[width=0.6\hsize,clip]{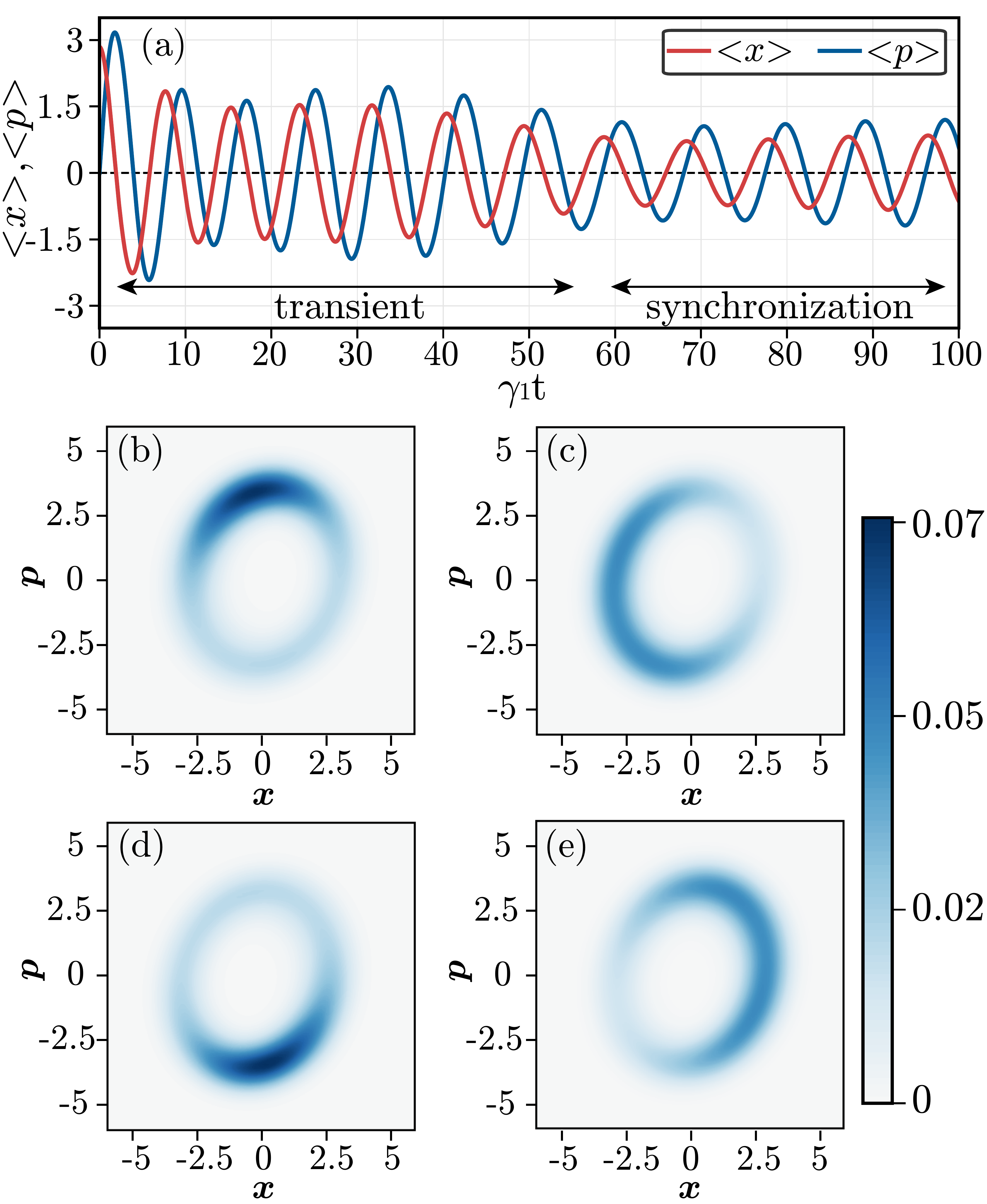}
		\caption{Synchronization of a quantum vdP oscillator subjected to 
			harmonic driving with periodic amplitude modulation.
			(a): Evolution of the averages $\la x \ra$ and $\la p \ra$ from a coherent-state initial condition.
			(b-e): Snapshots of the Wigner distributions in the periodic steady (cyclo-stationary) state at time $t =89.5$ (b), $91.8$ (c), $94.1$ (d), and $96.4$ (e), respectively.
			The parameters are given by $(\Delta, \gamma_{2}, \eta e^{i \theta}, E, )/\gamma_{1} = (0.8, 0.05, -0.1i, \sqrt{0.1})$ and $\Delta_{e} = 0.1$.
			}
		\label{fig_5}
	\end{center}
\end{figure}

In this case, we obtain a periodic (cyclo-stationary) solution of period $T_e = 2\pi / \omega_e$ instead of a stationary solution. 
As shown in Fig.~\ref{fig_5}(a), the quantum-mechanical averages $\langle {x} \rangle$ and $\langle {p} \rangle$ of the position and momentum operators ${x} = (a + a^{\dag}) / 2$ and ${p} = -i (a - a^{\dag}) / 2$ exhibit steady periodic dynamics after the initial transient.
Here, the initial condition is a coherent state $| \alpha_0(\phi = 0) \rangle$, where
${\bm \alpha}_0(\phi=0)$ is a point on the limit cycle with $\phi = 0$.
Figure~\ref{fig_5}(b)-(e) show snapshots of the Wigner distributions in the periodic
state, where the system evolves as (b) $\to$ (c) $\to$ (d) $\to$ (e) $\to$ (b)
(see Supplemental Video for the continuous evolution~\cite{supplement}).
The tendency toward synchronization can be clearly observed from the existence of the
dense region co-rotating with the external forcing.

We denote the quantum and approximated autocovariance functions 
at a given time $t_e$ ($0 \leq t_e < T_e$) of the steady state oscillation as $R_{qm,sc}^{t_e}(\tau)$,
where $R^{t_e}_{qm}(\tau)$ is calculated by using 
a density matrix $\rho_{qm}(t_e)$ at time $t_e$
and $R^{t_e}_{sc}(\tau)$ is calculated by using a phase distribution 
$P_{sc}(\phi, t_e)$ at time $t_e$,
respectively, in the steadily oscillating state.
Then we use the averaged power spectra
$\bar{S}_{qm, sc}(\omega)
= \int_{-\infty}^{\infty} d\tau e^{i\omega \tau} 
\int_{0}^{T_e} dt_e R_{qm,sc}^{t_e}(\tau)/T_e$
to evaluate the observed frequencies relative to the frequency of the amplitude modulation as $\bar{\omega}_{qm,sc} = \argmax_{\omega}\bar{S}_{qm, sc}(\omega)  - \omega_e$.
Figure~\ref{fig_4}(c) and (f) compare the averaged spectra $\bar{S}_{qm, sc}(\omega)$ and observed frequencies $\bar{\omega}_{qm, sc}$ obtained by direct numerical simulation of the original master equation and by the approximate phase equation, respectively.
It can be seen that the spectrum and observed frequency obtained from the original master equation are accurately reproduced by those obtained from the approximate phase equation.
Thus, by using the reduced phase equation, we can approximately reconstruct the spectrum and observed frequency of the original system also in this case.

%%%%%%%%%%%%%%%%%
%%% Seciton 4 %%%
%%%%%%%%%%%%%%%%%

\section{Concluding remarks}

We have developed a general framework of the phase reduction theory for quantum limit-cycle oscillators under the semiclassical approximation and confirmed its validity by analyzing synchronization dynamics of the quantum vdP model.
The proposed framework can approximately characterize the dynamics of a quantum nonlinear oscillator by using a simple classical phase equation, which would serve as a starting point for analyzing synchronization of quantum nonlinear oscillators under the semiclassical approximation.
Although we have only analyzed a single-oscillator problem with a single degree of freedom in this study, the developed framework can be directly extended to two or more quantum oscillators with weak coupling by using standard methods from the classical phase reduction theory. Analysis of large many-body systems and the study of their collective dynamics are of particular interest~\cite{lee2014entanglement, witthaut2017classical, ludwig2013quantum, lee2013quantum,davis2018dynamics}.

In this study, we have employed the P-representation for formulating the semiclassical phase reduction theory; however, other quasiprobability distributions 
can also be used for the formulation. Detailed comparisons of the results between different representations, including the positive-P representation which is necessary 
to treat negative-definite diffusion matrices~\cite{carmichael2007statistical}, will be discussed in our forthcoming studies.
Also, analysis on the genuine quantum signature of a quantum limit-cycle oscillator, which, for instance, can be measured by the negativity of a Wigner quasiprobability distribution \cite{weiss2017quantum,lorch2016genuine}, could be performed via an extended version of the developed phase reduction theory.

Recently, the phase reduction theory has been applied to control and optimization of synchronization dynamics in classical nonlinear oscillators ~\cite{harada2010optimal,zlotnik2012optimal,zlotnik2013optimal, pikovsky2015maximizing, watanabe2019optimization, monga2018synchronizing}. 
In classical dissipative systems, the phase reduction theory has already been used in technical applications of synchronization such as in the ring laser gyroscope~\cite{macek1963rotation,cresser1982quantum1, *cresser1982quantum2, *cresser1982quantum3}, phase-locked loop~\cite{best1984phase,pikovsky2001synchronization}, and Josephson voltage standard~\cite{josephson1962possible, shapiro1963josephson,pikovsky2001synchronization}.
The quantum version of these applications, as well as the recent demonstrations \cite{xu2015conditional,hamerly2015optical}, could be systematically investigated via the semiclassical phase reduction theory developed in the present study.
These subjects will also be discussed in our forthcoming studies.

\begin{acknowledgments}
This research was supported by the JSPS KAKENHI Grant Numbers JP16K13847, JP17H03279, 18K03471, and JP18H03287.
\end{acknowledgments}

\appendix

%%%%%%%%%%%%%%%%%
%%% Seciton A1 %%%
%%%%%%%%%%%%%%%%%

\section{Explicit form of $\bm{\beta}(\bm{\alpha})$}
\label{ap_1}

In this section, we derive an explicit expression of $\bm{\beta}(\bm{\alpha})$ in Eq.~(\ref{eq:ldv}).
The diffusion matrix of the FPE in Eq.~(\ref{eq:fpe}) in the complex representation is given by
\begin{align}
{\bm D}({\bm \alpha}) = \bm{\beta}(\bm{\alpha}) \bm{\beta}(\bm{\alpha})^T 
=
\begin{pmatrix}
D_{11}({\bm \alpha}) & D_{12}({\bm \alpha}) \\
D_{21}({\bm \alpha}) & D_{22}({\bm \alpha}) \\
\end{pmatrix}
\in {\mathbb C}^{2 \times 2},
\end{align}
where $D_{22}(\bm{\alpha}) = D_{11}^{*}(\bm{\alpha})$ and $D_{12}(\bm{\alpha}) = D_{21}(\bm{\alpha})$.
The non-diagonal element $ D_{12}(\bm{\alpha}) = D_{21}(\bm{\alpha})$ is real and positive, because it is a constant of cross diffusion described by $ \partial^2 P({\bm \alpha}, t) / \partial \alpha  \partial \alpha^*$ and it can be obtained as an absolute value of a complex variable.

We rewrite the FPE in Eq.~(\ref{eq:fpe}) corresponding to the SDE in Eq.~(\ref{eq:X}) in the real-valued representation, 
i.e., for the quasiprobability distribution $P({\bm X}, t)$ with $\bm{X} = (x, p)^T = (\mbox{Re}~\alpha, \mbox{Im}~\alpha)^T$, as
\begin{align}
\frac{\pa}{\pa t} P({\bm X}, t) = \left[ - \frac{\pa}{\pa {\bm X}} \{ {\bm F}({\bm X}) + \epsilon {\bm q}({\bm X}, t) \} + \frac{1}{2} \frac{\pa^2}{\pa {\bm X}^2} {\bm D}({\bm X}) \right] P({\bm X}, t),
\end{align}
where
\begin{align}
\frac{\partial}{\partial \alpha} = \frac{1}{2} \left( \frac{\partial}{\partial x} - i  \frac{\partial}{\partial p} \right),
\quad
\frac{\partial}{\partial \alpha^*} = \frac{1}{2} \left( \frac{\partial}{\partial x} + i  \frac{\partial}{\partial p} \right).
\end{align}
The real-valued diffusion matrix ${\bm D}({\bm X})$ in the above FPE and the complex-valued diffusion matrix ${\bm D}({\bm \alpha})$ are related as
\begin{align}
\label{eq:diffusion_real}
{\bm D}(\bm{X}) &= 
\frac{1}{4}
\begin{pmatrix}
1 & 1 \\ -i & i 
\end{pmatrix}
{\bm D}({\bm \alpha})
\begin{pmatrix}
1 & -i \\ 1 & i 
\end{pmatrix}
\cr
&=
\frac{1}{2} \left( \begin{matrix}
\mbox{Re}~D_{11}(\bm{\alpha}) + D_{12}(\bm{\alpha})  & \mbox{Im}~D_{11}(\bm{\alpha}) \\
\mbox{Im}~D_{11}(\bm{\alpha}) & - \mbox{Re}~D_{11}(\bm{\alpha}) + D_{12}(\bm{\alpha}) \\
\end{matrix} \right)
\in {\mathbb R}^{2 \times 2}
\end{align}
and
\begin{align}
\bm{D}(\bm{\alpha}) &= 
\begin{pmatrix}
1 & i \\ 1 & -i 
\end{pmatrix}
{\bm D}({\bm X})
\begin{pmatrix}
1 & 1 \\ i & -i
\end{pmatrix}.
\end{align}

By denoting the matrix components of ${\bm D}({\bm \alpha})$ in the polar representation as
$ D_{11}(\bm{\alpha}) = R_{11}(\bm{\alpha}) e^{i \chi({\bm \alpha})}$ and $ D_{12}(\bm{\alpha}) = R_{12}(\bm{\alpha})$, where $R_{11}({\bm \alpha}), R_{22}({\bm \alpha}) \geq 0$ and $\chi({\bm \alpha}) \in [0, 2\pi)$, the eigenvalues $\lambda_{\pm}(\bm{X})$ and 
eigenvectors $\bm{v}_{\pm}(\bm{X})$ 
of $ \bm{D}(\bm{X})$ can be expressed as
\begin{align}
\label{eq:eig_sys}
\lambda_{\pm}(\bm{X}) &= 
\frac{1}{2} \left( R_{12}(\bm{\alpha}) \pm  R_{11}(\bm{\alpha}) \right),
\cr
\bm{v}_{+}(\bm{X}) &= 
\left(
\begin{array}{c}
\cos \frac{\chi(\bm{\alpha})}{2} \\
\sin \frac{\chi(\bm{\alpha})}{2}
\end{array}
\right)
,
\quad
\bm{v}_{-}(\bm{X}) = 
\left(
\begin{array}{c}
\sin \frac{\chi(\bm{\alpha})}{2} \\
-\cos \frac{\chi(\bm{\alpha})}{2}
\end{array}
\right),
\end{align}
and ${\bm D}({\bm X})$ can be decomposed as 
\begin{align}
{\bm D}({\bm X}) =
\begin{pmatrix}
{\bm v}_+({\bm X}) & {\bm v}_-({\bm X})
\end{pmatrix}
\begin{pmatrix} \lambda_+({\bm X}) & 0 \\ 0 & \lambda_-({\bm X})
\end{pmatrix}
\begin{pmatrix}
{\bm v}_+({\bm X})^T \\
{\bm v}_-({\bm X})^T
\end{pmatrix}.
\end{align}
Thus, ${\bm G}({\bm X})$ is given by
\begin{align}
\bm{G}(\bm{X}) &= 
\begin{pmatrix} \bm{v}_{+}(\bm{X}) & \bm{v}_{-}(\bm{X}) \end{pmatrix}
\left( \begin{matrix}
\sqrt{\lambda_{+}(\bm{X})} & 0 \\
0 & \sqrt{\lambda_{-}(\bm{X})} \\
\end{matrix} \right)
\nonumber \\
& =
\left( \begin{matrix}
\sqrt{\frac{\left( R_{12}(\bm{\alpha}) +  R_{11}(\bm{\alpha}) \right)}{2}} 
\cos \frac{\chi(\bm{\alpha})}{2} & 
\sqrt{\frac{\left( R_{12}(\bm{\alpha}) -  R_{11}(\bm{\alpha}) \right)}{2}} 
\sin \frac{\chi(\bm{\alpha})}{2} \\
\sqrt{\frac{\left( R_{12}(\bm{\alpha}) +  R_{11}(\bm{\alpha}) \right)}{2}} 
\sin \frac{\chi(\bm{\alpha})}{2} & 
- \sqrt{\frac{\left( R_{12}(\bm{\alpha}) -  R_{11}(\bm{\alpha}) \right)}{2}}  
\cos \frac{\chi(\bm{\alpha})}{2} \\
\end{matrix} \right),
\end{align}
and $ \bm{\beta}(\bm{\alpha})$ is obtained from ${\bm G}({\bm X})$ as
\begin{align}
\label{eq:beta}
\bm{\beta}(\bm{\alpha}) &= 
\left( \begin{matrix}
1 & i \\
1 & -i \\
\end{matrix} \right)
\bm{G}(\bm{X})
=
\begin{pmatrix}
\sqrt{\frac{\left( R_{12}(\bm{\alpha}) +  R_{11}(\bm{\alpha}) \right)}{2}} e^{i \chi(\bm{\alpha})/ 2}
&
-i \sqrt{\frac{\left( R_{12}(\bm{\alpha}) -  R_{11}(\bm{\alpha}) \right)}{2}} e^{i \chi(\bm{\alpha}) / 2}
\\
\sqrt{\frac{\left( R_{12}(\bm{\alpha}) +  R_{11}(\bm{\alpha}) \right)}{2}} e^{- i \chi(\bm{\alpha}) / 2}
&
i \sqrt{\frac{\left( R_{12}(\bm{\alpha}) -  R_{11}(\bm{\alpha}) \right)}{2}} e^{- i \chi(\bm{\alpha}) / 2}
\end{pmatrix}.
\end{align}

The assumption in the main text that the diffusion matrix is always positive semidefinite along the limit cycle is equivalent to the assumption that $\lambda_{-}(\bm{X}_0(\phi)) \geq 0$, that is,
$R_{12}(\bm{\alpha}_0(\phi)) \geq  R_{11}(\bm{\alpha}_0(\phi)) $ is satisfied for all $\phi$,
because $\lambda_{+}(\bm{X})$ is always positive.
With this assumption, if the initial state is given in the form of Eq.~(\ref{eq:rho}), for instance, by a pure coherent state $\rho = \left| \alpha_{0}(\phi_{0}) \right\rangle \left\langle \alpha_{0}(\phi_{0}) \right|$ at a given phase point $\phi_{0}$ on the limit cycle, 
the state always remains in the two-dimensional phase space of the classical variables.

%%%%%%%%%%%%%%%%%
%%% Seciton A2 %%%
%%%%%%%%%%%%%%%%%

\section{Derivation of the phase equation}
\label{ap_2}

In this section, we give a detailed derivation of the phase equation in Eq.~(\ref{eq:dphi}). 
The asymptotic phase function $\Phi({\bm X}) : B \subset {\mathbb R}^{2 \times 1} \to [0, 2\pi)$ introduced in the main text satisfies
\begin{align}
\bm{F}({\bm X}) \cdot \nabla \Phi({\bm X}) = \omega
\end{align}
in the basin $B$ of the limit cycle, where $\nabla \Phi \in {\mathbb R}^{2 \times 1}$ indicates the gradient of $\Phi$ with respect to ${\bm X}$.
Using this $\Phi({\bm X})$, we define the phase $\phi$ of the oscillator state ${\bm X}$ as $\phi = {\Phi}({\bm X})$.
As long as ${\bm X}$ evolves in $B$, $\dot{\phi} = \dot{\Phi}({\bm X}) = \dot{\bm X} \cdot \nabla \Phi({\bm X}) = \bm{F}({\bm X}) \cdot \nabla \Phi({\bm X}) = \omega$ holds.
Recently, it has been shown that this phase function is closely related to an eigenfunction of the Koopman operator of the system $\dot{\bm X} = {\bm F}({\bm X})$ associated with the eigenvalue $i\omega$~\cite{mauroy2013isostables}.

When ${\bm X}$ obeys the Ito SDE in Eq.~(\ref{eq:X}), we obtain an Ito SDEs for the phase $\phi$ as
\begin{align}
d\phi
=& \left[ ( \nabla \Phi({\bm X}) ) \cdot ( {\bm F}({\bm X}) + \epsilon {\bm q}({\bm X}, t) ) + \frac{1}{2} \epsilon \mbox{Tr} \left\{ {\bm G}({\bm X})^T (\nabla^T \nabla \Phi({\bm X})) {\bm G}({\bm X}) \right\} \right] dt
\cr
& + \sqrt{\epsilon} ( \nabla \Phi({\bm X}) ) \cdot ( {\bm G}({\bm X}) d{\bm W} )
\cr
=& \left[ \omega + \epsilon ( \nabla \Phi({\bm X}) ) \cdot  {\bm q}({\bm X}, t) + \frac{1}{2} \epsilon \mbox{Tr} \left\{ {\bm G}({\bm X})^T (\nabla^T \nabla \Phi({\bm X})) {\bm G}({\bm X}) \right\} \right] dt
\cr
& + \sqrt{\epsilon} ( {\bm G}({\bm X})^T \nabla \Phi({\bm X}) ) \cdot  d{\bm W},
\end{align}
where the third term in the drift part arises from the change of the variables by the Ito formula and
$\nabla^T \nabla \Phi \in {\mathbb R}^{2 \times 2}$ represents the Hessian matrix of $\Phi({\bm X})$ with respect to ${\bm X}$.
This equation is still not closed in the phase variable $\phi$, because each term on the right-hand side depends on ${\bm X}$.

When the perturbation and quantum noise are weak, the deviation of the system state ${\bm X}$ from the limit cycle is small and of the order of $\mathcal{O}(\sqrt{\epsilon})$
because the limit cycle is exponentially stable and the system state is subjected to Gaussian-white noise. 
Thus, in the lowest-order approximation, we can approximate the state ${\bm X}$ by a state ${\bm X}_0(\phi)$ on the limit cycle as ${\bm X}(t) = {\bm X}_0(\phi(t)) +
\mathcal{O}(\sqrt{\epsilon})$. We then obtain an Ito SDE for the phase variable $\phi$,
\begin{align}
\label{eq:dphi0}
d \phi  &=  \left\{ \omega + \epsilon f(\phi, t) + \epsilon g(\phi) \right\} dt 
+ \sqrt{\epsilon} {\bm h}(\phi) \cdot d\bm{W},
\end{align}
which is correct up to $\mathcal{O}(\epsilon)$ in the drift term and up to $\mathcal{O}(\sqrt{\epsilon})$ in the noise intensity, where
\begin{align}
f(\phi, t) = \nabla\Phi({\bm{X}})|_{{\bm X} = {\bm X}_0(\phi) } \cdot {{\bm q}}({\bm{X}_0(\phi)}, t) \in {\mathbb R}
\end{align}
represents the effect of the perturbation on $\phi$, 
\begin{align}
{\bm h}(\phi) = {\bm G}({\bm{X}_0(\phi)})^T  \nabla\Phi({\bm{X}})|_{{\bm X} = {{\bm X}_0(\phi) }} \in {\mathbb R}^{2 \times 1}
\end{align}
represents the effect of the quantum noise on $\phi$, and
\begin{align}
g(\phi) = \frac{1}{2} \mbox{Tr} \left\{ {\bm G}({\bm X}_0(\phi))^T (\nabla^T \nabla \Phi|_{{\bm X} = {{\bm X}_0(\phi) }}) {\bm G}({\bm X}_0(\phi)) \right\}
\end{align}
represents a term arising from the change of the variables, respectively.

We denote the gradient vector (PSF) and Hessian matrix of the phase function
$\Phi({\bm X})$ evaluated at ${\bm X} = {\bm X}_0(\phi)$ on the limit cycle as 
$\bm{Z}( \phi)  = \nabla\Phi|_{ {\bm{X} }  = {\bm{X}}_{0}(\phi) }$
and
$\bm{Y}( \phi)  = \nabla^T \nabla \Phi|_{ {\bm{X} }  = {\bm{X}}_{0}(\phi) }$,
respectively.
The components of the PSF and Hessian matrix $\nabla^T \nabla \Phi|_{{\bm X} = {{\bm X}_0(\phi) }}$ are explicitly given by
\begin{align}
Z_i(\phi) = \left. \frac{ \partial \Phi({\bm X}) }{ \partial X_i } \right|_{{\bm X} = {\bm X}_0(\phi)},\quad
( \nabla^T \nabla \Phi|_{{ \bm X} = {\bm X}_0(\phi)} )_{ij} = \left. \frac{ \partial^2 \Phi({\bm X})}{ \partial X_i \partial X_j } \right|_{{\bm X} = {\bm X}_0(\phi)},
\end{align}
for $i, j = 1, 2$, respectively.

It is well known in the classical phase reduction theory~\cite{nakao2016phase, 
	ermentrout2010mathematical,ermentrout1996type,brown2004phase} that ${\bm Z}(\phi)$ 
is given by a $2\pi$-periodic solution to the following adjoint equation and normalization condition:
\begin{align}
\omega \frac{d}{d\phi} {\bm Z}(\phi) = - {\bm J}(\phi)^T {\bm Z}(\phi),
\quad
{\bm Z}(\phi) \cdot {\bm F}({\bm X}_0(\phi)) = \omega.
\end{align}
It is also known~\cite{suvak2010quadratic, takeshita2010higher} that the Hessian matrix ${\bm Y}(\phi)$ of the phase function, evaluated at ${\bm X} = {\bm X}_0(\phi)$ on the limit cycle, is given by a $2\pi$-periodic solution to a differential equation
\begin{align}
\omega \frac{d}{d\phi} {\bm Y}(\phi) = - {\bm J}(\phi)^T {\bm Y}(\phi) - {\bm Y}(\phi) {\bm J}(\phi) -  {\bm Z}(\phi) \circ {\bm K}(\phi),
\end{align}
which satisfies a constraint
\begin{align}
{\bm Z}(\phi) \cdot {\bm J}(\phi) {\bm F}({\bm X}_0(\phi)) + {\bm F}({\bm X}_0(\phi)) \cdot {\bm Y}(\phi) {\bm F}({\bm X}_0(\phi)) = 0.
\end{align}
In the above equations, ${\bm J}(\phi) \in {\mathbb R}^{2 \times 2}$ is a Jacobian matrix of ${\bm F}({\bm X})$ at ${\bm X} = {\bm X}_0(\phi)$ and ${\bm K}(\phi) \in {\mathbb R}^{2 \times 2 \times 2}$ is a third order tensor, respectively, whose components are given by
\begin{align}
J(\theta)_{ij} = \left. \frac{\partial F_i}{\partial X_j} \right|_{{\bm X} = {\bm X}_0(\theta)},
\quad
K(\theta)_{ijk} = \left. \frac{\partial^2 F_i}{\partial X_j \partial X_k} \right|_{{\bm X} = {\bm X}_0(\theta)},
\end{align}
and the matrix components of the product ${\bm Z}(\phi) \circ {\bm K}(\phi) \in {\mathbb R}^{2 \times 2}$ are given by
\begin{align}
[ {\bm Z}(\phi) \circ {\bm K}(\phi) ]_{j, k} = \sum_{i=1}^2 Z_i(\phi) K_{ijk}(\phi)
\end{align}
for $i, j, k = 1, 2$.

Thus, when the noise and perturbations are sufficiently weak, we obtain an approximate Ito SDE for the phase variable as
\begin{align}
d\phi = \{ \omega + \epsilon f(\phi, t) + \epsilon g(\phi) \}dt + \sqrt{\epsilon} {\bm h}(\phi) \cdot d\bm{W}
\label{eq:dphi2}
\end{align}
at the lowest order, which corresponds to Eq.~(\ref{eq:dphi}) in the main text. 
It can be shown that the amplitude effect does not enter the phase dynamics at the lowest order \cite{shirasaka2017phase}.
As Eq.~(\ref{eq:dphi2}) is an Ito SDE, using the property of the Wiener process, the noise term can be rewritten as
\begin{align}
\sqrt{\epsilon} {\bm h}(\phi) \cdot d\bm{W} = \sqrt{\epsilon} h(\phi) dW,
\end{align}
where $h(\phi) = \sqrt{ \sum_{i=1}^2 ( {\bm h}(\phi) )_i^2 }$ and $W(t)$ is a one-dimensional Wiener process.

The errors in the evolution of the phase variable resulting from the lowest-order approximation above are $\mathcal{O}(\epsilon^2)$ in the drift term and $\mathcal{O}(\epsilon)$ in the noise intensity, respectively.
Therefore, the error in the mean of $\phi$ from the true value grows with time as $\mathcal{O}(\epsilon^2 t)$, and the error in the variance of $\phi$ grows as $\mathcal{O}(\epsilon^2 t)$.
Thus, these errors in the phase dynamics remain $\mathcal{O}(\epsilon)$ up to $t = \mathcal{O}(1/\epsilon)$.

\section{Averaged phase equation}
\label{ap_3}

In this section, we derive the averaged phase equation, Eq.~(\ref{eq:averagedphase}), by using the near-identity transform.
Although Eq.~(\ref{eq:dphi}) is a correct phase equation for the phase $\phi$ in the lowest-order approximation, it has an additional function $g(\phi)$ in the drift term, which adds tiny periodic fluctuations to the deterministic part. By further introducing a new phase $\psi$ that is only slightly different from $\phi$, we can eliminate this term and obtain a simpler SDE,
\begin{align}
\label{eq:dphinit}
d\psi = \{ \tilde{\omega} + \epsilon f(\psi, t) \} dt
+ \sqrt{\epsilon} h(\psi) dW,
\end{align}
where $f(\psi, t) = \bm{Z}( \psi ) \cdot {\bm q}(\psi, t)$, $W(t)$ is a one-dimensional Wiener process, and $h(\psi)$ is a $2\pi$-periodic function of $\psi$.
Here, the new phase $\psi$ is defined from $\phi$ by a near-identity transform as $\phi = \psi + \epsilon n(\psi)$, where $n(\psi)$ is a $2\pi$-periodic function with $n(0) = 0$.
Using this transformation, the additional term $g(\phi)$ in Eq.~(\ref{eq:dphi}) can be renormalized into the frequency term as
\begin{align}
\tilde{\omega} = \omega + \frac{\epsilon}{2\pi} \int_0^{2\pi} g(\psi') d\psi',
\label{eq:effectivefreq_ap}
\end{align}
where $\tilde{\omega}$ is the effective frequency of the system.
As $\epsilon$ is assumed to be sufficiently small, the transformation between the two variables $\phi$ and $\psi$ is invertible. 
Thus, the qualitative properties of the dynamics predicted by the two-phase equations, such as whether synchronization occurs or not, are invariant. In the classical phase-reduction theory, the $\mathcal{O}(\epsilon)$ difference between the phase variables due to the near-identity transformation or averaging is often neglected and both phases are considered to be the same.
Below, we derive the simplified phase equation in Eq.~(\ref{eq:dphinit}) from the original phase equation, Eq.~(\ref{eq:dphi}) or (\ref{eq:dphi2}), by using the near-identity transform~\cite{sanders1985averaging}.

%%%%%%%%%%%%%%%%%
%%% Seciton A4 %%%
%%%%%%%%%%%%%%%%%

In Eq.~(\ref{eq:dphi2}), the function $g(\phi)$ contains the Hessian matrix ${\bm Y}(\phi)$ of $\Phi({\bm X})$ on the limit cycle, which is typically not included in the phase equation for classical limit-cycle oscillators and gives a tiny but complex periodic contribution to the phase dynamics.
To eliminate this term, we renormalize it into the frequency term. 
For this purpose, we consider a near-identity transform from the original phase $\phi$ to a new phase $\psi$, 
\begin{align}
\phi = \psi + \epsilon n(\psi),
\end{align}
where the transformation function $n(\psi)$ is a smooth $2\pi$-periodic function of $\psi$ satisfying $n(0) = 0$, and assume that $\psi$ obeys an Ito SDE of the form
\begin{align}
d\psi
&=  \{ \omega + \epsilon \Omega + \epsilon f(\psi, t) \} dt + \sqrt{\epsilon} h(\psi) dW
\label{eq:dpsi0}
\end{align}
in the lowest-order approximation, which does not contain a term corresponding to $g(\phi)$ but has a small shift $\epsilon \Omega$ in the frequency.
From this SDE, we obtain an Ito SDE for $\phi$ by using the Ito formula as
\begin{align}
d\phi 
&= \left[ \frac{\partial \phi}{\partial \psi} \{ \omega + \epsilon \Omega + \epsilon f(\psi, t) \} + \frac{1}{2} \epsilon
h(\psi)^2 \frac{\partial^2 \phi}{\partial \psi^2} \right] dt + \sqrt{\epsilon} \frac{\partial \phi}{\partial \psi} h(\psi) dW
\cr
&= \left[ ( 1 + \epsilon n'(\psi) ) \{ \omega + \epsilon \Omega + \epsilon f(\psi, t) \} + \frac{1}{2} \epsilon
h(\psi)^2 ( \epsilon n''(\psi) ) \right] dt + \sqrt{\epsilon} ( 1 + \epsilon n'(\psi) ) h(\psi) dW
\cr
&\approx
\left[\omega + \epsilon f(\psi, t) + \epsilon \Omega + \epsilon \omega n'(\psi) \right] dt + \sqrt{\epsilon} h(\psi) dW,
\end{align}
where we omitted the tiny terms of $\mathcal{O}(\epsilon^2)$ in the drift term and $\mathcal{O}(\epsilon^{3/2})$ in the noise intensity. 
The replacement of $\phi$ by $\psi$ in the functions $f$ and $h$ also results in errors of $\mathcal{O}(\epsilon^2)$ and $\mathcal{O}(\epsilon^{3/2})$ 
in the drift term and noise intensity, respectively, which can also be neglected.

The above equation coincides with the original Eq.~(\ref{eq:dphi2}) if $n(\psi)$ satisfies
\begin{align}
\Omega + \omega n'(\psi) = g(\phi).
\end{align}
As $g(\phi) = g(\psi) + \mathcal{O}(\epsilon)$, the equation for $n(\psi)$ is obtained at the lowest order as
\begin{align}
\frac{d}{d\psi} n(\psi) = g(\psi) - \Omega,
\end{align}
which gives
\begin{align}
n(\psi) = \int_0^\psi d\psi' \left[ g(\psi') - \Omega \right],
\end{align}
where $n(0) = 0$ is used. Moreover, as $n(\psi)$ is $2\pi$-periodic, $n(2\pi) = n(0) = 0$ should hold, which determines the frequency shift $\Omega$ as
\begin{align}
\epsilon \Omega = \frac{\epsilon}{2\pi} \int_0^{2\pi} d\psi' g(\psi').
\label{eq:shift}
\end{align}
Thus, by introducing the near-identity transform, we obtain an averaged phase equation
\begin{align}
d\psi = \{ \tilde{\omega} + \epsilon f(\psi, t) \} dt + \sqrt{\epsilon} h(\psi) dW,
\label{eq:dpsi2}
\end{align}
where $\tilde{\omega} = \omega + \epsilon \Omega$ is a renormalized, effective frequency. This corresponds to Eq.~(\ref{eq:dphinit}).
The orders of errors caused by the above near-identity transformation are $\mathcal{O}(\epsilon^2)$ in the drift term and $\mathcal{O}(\epsilon^{3/2})$ in the noise intensity.
Therefore, the phase equations in Eq.~(\ref{eq:dphi2}) and ~(\ref{eq:dpsi2}) are equally correct in the lowest-order approximation and valid up to $t = \mathcal{O}(1/\epsilon)$.

The frequency shift $\epsilon \Omega$ can be evaluated by numerically calculating the Hessian matrix of $\Phi({\bm X})$ in $g(\psi)$ and integrating Eq.~(\ref{eq:shift}), or alternatively by measuring $\tilde{\omega}$ by numerically evolving the SDE in Eq.~(\ref{eq:ldv}) or Eq.~(\ref{eq:X}) without perturbations.
In the examples used in the main text, the frequency shift $\epsilon \Omega$ is zero in the case of Eq.~(\ref{eq:qvdp_phi}) with the symmetric limit cycle with weak squeezing, and takes a tiny value in the case with strong squeezing.
In other applications, for example, in the analysis of coupled identical limit-cycle oscillators without external forcing, the precise value of $\tilde{\omega}$ may not be required (only the frequency difference matters). In such cases, one may simply assume $\tilde{\omega} \approx \omega$ and avoid the calculation of $\epsilon \Omega$.

%%%%%%%%%%%%%%%%%
%%% Seciton A5 %%%
%%%%%%%%%%%%%%%%%

\section{Phase-space representation of a quantum vdP oscillator with harmonic driving and squeezing}
\label{ap_5}

\subsection{Weak squeezing}

Here, we derive a phase equation for a quantum vdP oscillator with harmonic driving and  squeezing.
In the case of weak squeezing with $\delta = \epsilon$, the rescaled system Hamiltonian and the perturbation Hamiltonian are given by
\begin{align}
\label{eq:qvdp_H1}
H = - \Delta' a'^{\dag}a',
\quad
\epsilon {\tilde{H}} = \epsilon \left\{ i  E' (a' - a'^{\dag})
+ i \eta' ( a'^2 e^{-i \theta} - a'^{\dag 2} e^{ i \theta}) \right\},
\end{align}
respectively, where the squeezing term is included in the perturbation.
The functions $\bm{A}(\bm{\alpha'})$, $\epsilon \bm{A}'(\bm{\alpha'})$, and $\epsilon \bm{D}(\bm{\alpha'})$ in the quantum FPE are calculated as
\begin{align}
\label{eq:qvdp_A1}
\bm{A}(\bm{\alpha'})
&= \left( \begin{matrix}
\left(\frac{1}{2} + i \Delta' \right) \alpha'   
- \gamma'_{2}\alpha'^{*} \alpha'^{2} 
\\
\left(\frac{1}{2} - i \Delta' \right) \alpha'^{*}   
- \gamma'_{2} \alpha'\alpha'^{*2} 
\\
\end{matrix} \right),
\quad
\epsilon \bm{A'}(\bm{\alpha'})
=
\epsilon 
\left( \begin{matrix}
- E' - 2 \eta' e^{i \theta}\alpha'^{*} \\
- E' - 2 \eta' e^{- i \theta}\alpha'\\
\end{matrix} \right),
\end{align}
and
\begin{align}
\label{eq:qvdp_D1}
\epsilon \bm{D}(\bm{\alpha'}) &= 
\epsilon 
\left( \begin{matrix}
- \gamma'_{2}\alpha'^{2} & 1  \\
1 & - \gamma'_{2}\alpha'^{*2} \\
\end{matrix} \right),
\end{align}
where the tiny terms of $\mathcal{O}(\epsilon^2)$ in $\epsilon {\bm D}({\bm \alpha}')$ are dropped.
The explicit form of $\bm{\beta}(\bm{\alpha}')$ given by Eq.~(\ref{eq:beta}) 
can be obtained from Eq.~(\ref{eq:qvdp_D1}) as
\begin{align}
{\bm \beta}({\bm \alpha}')
=
\left( \begin{matrix}
i \sqrt{\frac{1 + \gamma'_{2} R'^2}{2}} e^{i \delta'}   & \sqrt{\frac{1 - \gamma'_{2} R'^2}{2}} e^{i \delta'} \\
-i \sqrt{\frac{1 + \gamma'_{2} R'^2}{2}} e^{-i \delta'} & \sqrt{\frac{1 - \gamma'_{2} R'^2}{2}} e^{-i \delta'} \\
\end{matrix} \right),
\end{align}
where the modulus $R'$ and argument $\delta'$ of $\alpha'$ is introduced as $\alpha' = R' e^{i \delta'}$.
In the real-valued representation with ${\bm X} = (x', p')^T = ( \mbox{Re}\ \alpha, \mbox{Im}\ \alpha)^T$, the functions ${\bm F}({\bm X})$, $\epsilon {\bm q}({\bm X})$, and $\sqrt{\epsilon} {\bm G}({\bm X})$ are given by
\begin{align}
{\bm F}({\bm X})
=
\left( \begin{matrix}
\frac{1}{2}x'  - \Delta' p'  
- \gamma'_{2} x' (x'^{2} + p'^{2}) 
\\
\frac{1}{2} p'  + \Delta' x'  
- \gamma'_{2} p' (x'^{2} + p'^{2}) 
\\
\end{matrix} \right),
\;\;
\epsilon {\bm q}({\bm X})
=
\epsilon \left( \begin{matrix}
- E' - 2 \eta' ( x' \cos \theta + p' \sin \theta) \\
2 \eta' ( p' \cos \theta - x' \sin \theta) \\
\end{matrix} \right),
\end{align}
and
\begin{align}
\sqrt{\epsilon} {\bm G}({\bm X})
=
\sqrt{\epsilon}
\left( \begin{matrix}
- \sqrt{\frac{1 + \gamma'_{2} R'^2}{2}} \sin \delta' & \sqrt{\frac{1 - \gamma'_{2} R'^2}{2}} \cos \delta' \\
\sqrt{\frac{1 + \gamma'_{2} R'^2}{2}} \cos \delta' & \sqrt{\frac{1 - \gamma'_{2} R'^2}{2}}\sin \delta' \\
\end{matrix} \right),
\end{align}
respectively.

As discussed in the main text, the deterministic part of this equation, $\dot{\bm X} ={\bm F}({\bm X})$, is a normal form of the supercritical Hopf bifurcation, 
also known as the Stuart-Landau oscillator, and it is analytically solvable.
The limit cycle of this system in the classical limit can be obtained as ${\bm X}_0(\phi) = \sqrt{\frac{1}{2\gamma'_{2}}} (\cos \phi, \sin \phi)^T$ with $\phi = \omega t$, or ${\bm \alpha}_0'(\phi) = \sqrt{\frac{1}{2\gamma'_{2}}} ( e^{i \phi}, e^{-i \phi} )^T$ in the complex-valued representation, where the natural frequency is given by $\omega = \Delta'$, and the frequency shift $\epsilon \Omega$ vanishes.
From Eq.~(\ref{eq:qvdp_D1}), the eigenvalues of matrix ${\bm D}({\bm \alpha})$ can be calculated as
\begin{align}
\lambda_{\pm}(\bm{X}) &= 
\frac{1}{2} \left\{ R_{12}(\bm{\alpha'}) \pm  R_{11}(\bm{\alpha'}) \right\}
= 
\frac{1}{2} \left( 1 \pm \gamma_2' | \alpha'|^2 \right).
\end{align}
By plugging the limit-cycle solution ${\bm X}_0(\phi)$ into this equation, it can be seen that $\lambda_{-}(\bm{X}_{0}(\phi)) = \frac{1}{4} > 0$ is satisfied for any $\phi$ on the limit cycle and the diffusion matrix is always positive semidefinite along the limit cycle,
because the magnitudes of the squeezing and nonlinear damping,
which can cause negative diffusion, are assumed to be sufficiently small. 

\subsection{Strong squeezing}

In the case of strong squeezing with $\delta = 1$, the rescaled system Hamiltonian and the perturbation Hamiltonian are given by
\begin{align}
H = - \Delta' a'^{\dag}a' + i \eta' ( a'^2 e^{-i \theta} - a'^{\dag 2} e^{ i \theta} ),
\quad
\epsilon {\tilde{H}} = i \epsilon E' (a' - a'^{\dag}),
\end{align}
respectively, where the squeezing term is included in the system Hamiltonian. The functions $\bm{A}(\bm{\alpha'})$, $\bm{A}'(\bm{\alpha'})$, and $\bm{D}(\bm{\alpha'})$ in the phase-space representation are given by
\begin{align}
\bm{A}(\bm{\alpha'})
&= \left( \begin{matrix}
\left(\frac{1}{2} + i \Delta' \right) \alpha'   
- \gamma'_{2}\alpha'^{*} \alpha'^{2} 
- 2 \eta' e^{i \theta}\alpha'^{*} 
\\
\left(\frac{1}{2} - i \Delta' \right) \alpha'^{*}   
- \gamma'_{2} \alpha'\alpha'^{*2} 
- 2 \eta' e^{- i \theta}\alpha'
\\
\end{matrix} \right),
\quad
\epsilon \bm{A'}(\bm{\alpha'})
=
\epsilon \left( \begin{matrix}
- E' \\
- E' \\
\end{matrix} \right),
\end{align}
and
\begin{align}
\epsilon \bm{D}(\bm{\alpha'}) &= 
\epsilon \left( \begin{matrix}
- ( \gamma'_{2}\alpha'^{2}  + 2 \eta' e^{ i \theta} )   & 1  \\
1 & - ( \gamma'_{2}\alpha'^{*2}  + 2 \eta' e^{ - i \theta}) \\
\end{matrix} \right).
\end{align}
The explicit form of $\bm{\beta}(\bm{\alpha}')$ in this case is given by 
\begin{align}
\bm{\beta}(\bm{\alpha'}) 
&=
\begin{pmatrix}
\sqrt{\frac{\left( 1 +  R'_{2} \right)}{2}} e^{i \chi'_2 / 2}
&
-i \sqrt{\frac{\left(  1 -  R'_{2} \right)}{2}} e^{i \chi'_2 / 2}
\\
\sqrt{\frac{\left(  1 +  R'_{2} ) \right)}{2}} e^{- i \chi'_2 / 2}
&
i \sqrt{\frac{\left(  1 -  R'_{2} \right)}{2}} e^{- i \chi'_2 / 2}
\end{pmatrix},
\end{align}
where $R'_{2}e^{i \chi'_2} = -(\gamma'_2 \alpha'^2 + 2 \eta' e^{i\theta})$.
In the real-valued representation with ${\bm X} = (x', p')^T = ( \mbox{Re}\ \alpha, \mbox{Im}\ \alpha)^T$, the functions ${\bm F}({\bm X})$, $\epsilon {\bm q}({\bm X})$, and $\sqrt{\epsilon} {\bm G}({\bm X})$ are given by
\begin{align}
{\bm F}({\bm X})
=
\left( \begin{matrix}
\frac{1}{2}x'  - \Delta' p'  
- \gamma'_{2} x' (x'^{2} + p'^{2}) 
 - 2  \eta' ( x' \cos \theta + p' \sin \theta)
\\
\frac{1}{2} p'  + \Delta' x'  
- \gamma'_{2} p' (x'^{2} + p'^{2}) 
+ 2  \eta' ( p' \cos \theta - x' \sin \theta) 
\\
\end{matrix} \right),
\;\;
\epsilon {\bm q}({\bm X})
=
\epsilon \left( \begin{matrix}
- E' \\
0\\
\end{matrix} \right),
\end{align}
and
\begin{align}
\bm{G}(\bm{X}) & =
\left( \begin{matrix}
\sqrt{\frac{\left( 1 +  R'_{2} \right)}{2}} 
\cos \frac{\chi'_2}{2} & 
\sqrt{\frac{\left( 1 -  R'_{2} \right)}{2}} 
\sin \frac{\chi'_2}{2} \\
\sqrt{\frac{\left( 1 +  R'_{2} \right)}{2}} 
\sin \frac{\chi'_2}{2} & 
- \sqrt{\frac{\left( 1 -  R'_{2} \right)}{2}}  
\cos \frac{\chi'_2}{2} \\
\end{matrix} \right),
\end{align}
respectively.  

The deterministic part ${\bm F}({\bm X})$ gives an asymmetric limit cycle when $\eta > 0$, which is difficult to solve analytically.  However, we can still obtain the limit cycle ${\bm X}_0(\phi)$ numerically and use it to evaluate the PSF ${\bm Z}(\phi)$, Hessian matrix ${\bm Y}(\phi)$, and the noise intensity ${\bm G}(\phi)$, and use these quantities in the phase equation.
The PSF ${\bm Z}(\phi)$ can be numerically calculated by the adjoint method, and the Hessian matrix ${\bm Y}(\phi)$ can be calculated by using a shooting-type numerical algorithm.

When the squeezing is too strong, the diffusion matrix can generally be negative definite on the limit cycle. We choose parameter settings where the diffusion matrix is always positive semidefinite along the limit cycle in the main text.

\bibliographystyle{apsrev4-1}
%\bibliography{reference}

\end{document}